\documentclass{article}
\usepackage{preprint,times}

\usepackage[utf8]{inputenc}
\usepackage[T1]{fontenc}
\usepackage{hyperref}
\usepackage{url}
\usepackage{booktabs}
\usepackage{amsfonts}
\usepackage{nicefrac}
\usepackage{microtype}
\usepackage{xcolor}
\usepackage{xspace}
\usepackage{graphicx}
\usepackage{amsmath}
\usepackage{makecell}
\usepackage{multirow}
\usepackage{enumitem}
\usepackage{listings}
\usepackage{longtable}
\usepackage{needspace}
\setlist[itemize]{
	leftmargin=1em,
	itemindent=0.5em,
	labelsep=0.3em
}
\setlist[enumerate]{
	leftmargin=1em,
	itemindent=0.5em,
	labelsep=0.3em
}

\newcommand{\nruns}{$3$}

\newtheorem{example}{Example}[section]
\newtheorem{definition}{Definition}[section]

\newcommand{\bench}{\textsc{DatalogBench}\xspace}
\newcommand{\repourl}{\url{https://github.com/ZJU-PL/DatalogBench}}
\newcommand{\horn}{\ensuremath{\mathrel{\mathsf{\mathord{:}\mathord{-}}}}}

\title{\bench: Evaluating Large Language Models on Text-to-Datalog Synthesis}

\author{
	Yuan Li$^{1}$ \quad Hanyun Jiang$^{1}$ \quad Guowei Tian$^{1}$ \quad Chengpeng Wang$^{2}$ \quad Peisen Yao$^{1}$\\
	$^1$The State Key Laboratory of Blockchain and Data Security, Zhejiang University\\
	$^2$National University of Singapore\\
	\texttt{\{ly.liyuan,jhanyun,guow\_tian,pyaoaa\}@zju.edu.cn} \\
	\texttt{wang-chengpeng@nus.edu.sg} \\
}

\begin{document}

\maketitle

\begin{abstract}
	Datalog underpins reasoning tasks such as program analysis, but its programs are
	hard to write. Existing synthesizers automate this task but require users
	to state their intent as input--output examples. Large language models (LLMs)
	suggest a more natural route, text-to-Datalog synthesis from a natural-language
	question, yet how well they do so has not been systematically evaluated. We
	present \bench, a benchmark of $136$ text-to-Datalog synthesis tasks curated from existing
	Datalog-based artifacts. Synthesized programs are graded by execution on held-out inputs against an
	oracle validated by mutation analysis. Across six LLMs and four prompting
	configurations, exact match peaks at $75.7\%$; an input--output example raises
	it by about six points, while relation descriptions add little. Under
	direct prompting, most failures occur at compile time, typically because a model
	invents auxiliary predicates that it never declares or types consistently. Two coding
	agents reach up to $86.8\%$ and eliminate nearly all such failures, leaving mostly
	semantic errors concentrated in recursive tasks. \bench thus identifies
	recursive reasoning and decomposition as open challenges for current LLMs and
	agents, and offers a reliable, execution-grounded measure of both.
\end{abstract}

\section{Introduction}

Datalog~\citep{abiteboul1995foundations} is a declarative logic programming
language over the function-free fragment of Horn clauses, with a semantics given
by least-fixed-point evaluation. Expressive recursion and predictable semantics
have made it a workhorse in domains where clarity and global reasoning both
matter, such as static program
analysis~\citep{bravenboer2009strictly,whaley2004cloning,scholz2016fast}, network
verification~\citep{loo2006declarative}, distributed systems~\citep{alvaro2010boom},
and large-scale
analytics~\citep{halperin2014demonstration,seo2013socialite,shkapsky2016big}.
In industrial code analysis, CodeFuse-Query~\citep{xie2024codefuse} uses the
Datalog-based G\"odel language to express queries.
However, Datalog programs are hard to write, since developers must reason at
once about recursion, joins, and stratified negation.

This difficulty has motivated work on \emph{Datalog program
synthesis}~\citep{albarghouthi2017constraint,si2018syntax,si2019synthesizing,bembenek2023smt2asp,raghothaman2020provenance},
which offers strong guarantees for small programs. Its interface, however, is not
the one a user would choose. These systems, and the benchmarks built around
them, take the specification as \emph{input--output examples}: tables of facts
that the target program must derive and must not derive, often together with a
template or language bias that bounds the search~\citep{si2018syntax}. Stating
intent that way is neither natural nor easy. It presumes that the user can
already produce instances of the relation they are trying to define, and producing them is
most of the work on exactly the tasks where a synthesizer would help.

Natural language offers a more accessible interface, and large language models
(LLMs) make it plausible: given a question and a relational schema, a model may
generate the Datalog program directly. LogicLoc~\citep{xu2026logicloc} already
uses LLM-generated Datalog for code localization, but synthesis reliability has
not been systematically evaluated. Studies of code generation have repeatedly shown that
executable programs may nevertheless be semantically
incorrect~\citep{chen2021evaluating,liu2023isurcode,jimenez2024swebench}, and Datalog
programs require not only syntactic validity but precise declarative semantics. This raises a fundamental question:
\emph{Can probabilistic language models reliably synthesize programs that satisfy Datalog's strict logical semantics?}

Evaluating text-to-Datalog synthesis requires accounting for three properties of
the language (\S\ref{sec:prelim}).
First, correctness is judged on the least fixed point a program computes,
so a benchmark must contain tasks with recursive reference programs and grade
each program by executing it. Second, many targets decompose into auxiliary
predicates that the schema does not provide, so a benchmark must leave this
\emph{predicate invention} to the model. Third, Datalog programs are checked statically,
and a single mis-bound variable, type clash, or stratification violation rejects
an otherwise plausible program, so a benchmark must report compile failures separately from semantic errors.

We introduce \bench, a benchmark of $136$ text-to-Datalog synthesis tasks curated
from existing Datalog-based artifacts and spanning four reasoning domains:
program analysis, graph analytics, formal reasoning, and knowledge discovery.
Each task pairs a natural-language question with relation signatures, optional
relation descriptions, one demonstration instance, and a disjoint set of held-out
evaluation variants on which an execution oracle grades the synthesized program.
Since Datalog has no standardized surface syntax, \bench targets the dialect
of Souffl{\'e}~\citep{jordan2016souffle}, the engine underlying the artifacts our
tasks are drawn from, and every result we report holds for that dialect. We
evaluate six LLMs, each under four prompting configurations that vary the
relation schema and whether an input--output example is given, as well as two coding agents.

With direct prompting, exact match tops out at $75.7\%$; an input--output example
helps most models, while relation descriptions add little. The coding agents solve up to $86.8\%$, well above the same
models prompted directly. Most programs that fail under direct prompting are
rejected by the compiler, usually because an invented auxiliary predicate is
never declared or is typed inconsistently. Supplying the missing declarations
without changing any rule recovers part of the loss.
With the coding agents, compile failures nearly disappear, and the remaining
errors are mostly semantic and lie mainly in recursive tasks. The agents narrow
the gaps on recursive tasks and tasks whose reference programs use auxiliary predicates, but
the $47$ tasks whose references have both properties remain substantially harder than the $46$ whose references have neither.
Recursive reasoning and decomposition thus remain open challenges for current LLMs and agents.

In summary, we make the following contributions:
\begin{itemize}
	\item We present \bench, a benchmark of $136$ text-to-Datalog synthesis tasks curated from existing Datalog-based artifacts across four reasoning domains: program analysis, graph analytics, formal reasoning, and knowledge discovery.
	\item We provide held-out evaluation variants and a mutation-based audit of their discrimination, and report compile pass separately from exact output agreement.
	\item We evaluate six LLMs under four prompting configurations and two coding agents on \bench, and analyze their failures by error type and by task structure.
\end{itemize}

\section{Related Work}

\noindent \textbf{Synthesizing Datalog programs.}
Datalog program synthesis has largely relied on search and solving, using constraints, syntax-guided
search, numerical relaxation, solver-based rule selection, provenance-guided
refinement, or example-guided and evolutionary
search~\citep{albarghouthi2017constraint,si2018syntax,si2019synthesizing,
bembenek2023smt2asp,raghothaman2020provenance,thakkar2021example,mendelson2021gensynth}.
A closely related line of work, inductive logic programming (ILP), learns logic
programs, Datalog among them, from examples and background
knowledge~\citep{cropper2021popper,cropper2022ilp30}, and has long studied predicate
invention~\citep{muggleton2015meta}, recently together
with negation~\citep{cerna2024generalisation}. All of these take input--output examples
as the specification, typically with a language bias or rule templates.
LLMs have entered the picture either to steer a symbolic
search~\citep{barke2024hysynth} or even to write Datalog
queries inside a code-localization agent that validates them with a parser and
mutation-based feedback~\citep{xu2026logicloc}. In the latter, the Datalog program
is a means to localization and is scored by localization accuracy. To our knowledge,
no existing benchmark checks whether the Datalog programs that LLMs write from
natural-language questions compute the intended relations.

\smallskip
\noindent \textbf{Evaluating LLMs on code generation.}
HumanEval~\citep{chen2021evaluating} established function-level evaluation from
natural-language specifications, and EvalPlus~\citep{liu2023isurcode} showed that
weak tests overestimate correctness. Later benchmarks move toward realistic
software settings---repository- and class-level
synthesis~\citep{li2024deveval,yu2024codereval,du2024evaluating}, feature-level
development~\citep{li2025feabench}, issue
resolution~\citep{jimenez2024swebench,zhang2025swebenchgoeslive}, and
contamination-resistant, continuously updated evaluation~\citep{jain2025livecodebench}---but all target
imperative languages, where an incorrect program usually still runs and the strength
of the tests decides what the benchmark can detect. We take the lesson of
EvalPlus as a requirement and validate our oracle by mutation analysis
(\S\ref{sec:validation}). In Datalog, by contrast, many incorrect programs
never run: the compiler rejects a program that uses an undeclared relation or
violates typing, safety, or stratification, so an evaluation must report compile
failures separately from semantic errors (\S\ref{sec:failure}).

\smallskip
\noindent \textbf{Text-to-SQL benchmarks.}
WikiSQL~\citep{zhong2017seq2sql}, Spider~\citep{yu2018spider},
BIRD~\citep{li2023can}, and Spider 2.0~\citep{lei2025spider2} evaluate
SQL generation from natural language under increasingly realistic
schemas, emphasizing schema grounding, joins, aggregation, and value usage.
However, the reported construct inventories of Spider and BIRD do not list recursive
queries, while Spider 2.0 has no separate recursion-specific evaluation.
Although SQL supports intermediate results through subqueries or named CTEs,
these benchmarks do not report performance by auxiliary-predicate use in reference
programs. Appendix~\ref{app:benchmark-comparison} compares these benchmarks and
Datalog synthesis settings by specification, structural evaluation, and grading.

\section{Preliminaries}
\label{sec:prelim}

\noindent\textbf{Datalog.}
A Datalog program $P$ is a finite set of rules $A \horn B_1, \dots, B_n$, whose head $A$ and
body literals $B_i$ apply a predicate to a tuple of constants and variables. Predicates split into two classes:
\emph{extensional database} (EDB) predicates, defined by ground facts and supplied as
input, and \emph{intensional database} (IDB) predicates, defined by rules. The IDB holds both
the output relations a task is graded on and the auxiliary predicates the program
introduces to compute them. Given input facts $I$, the program's output
$P(I)$ is the least fixed point reached by applying the rules until nothing new
is derived. A program is \emph{recursive} when its predicate dependency graph, which consists of
edges from each body predicate to the head predicate of its rule, contains a cycle (through one predicate or several).
Pure Datalog restricts rules to Horn clauses, in which every $B_i$ is a positive atom. Like most practical dialects,
Souffl{\'e} also allows negated body literals and aggregation.
Rules may use negation only when it is \emph{stratified}: no cycle in the dependency graph
passes through a negated literal, so every negated predicate is fully computed before it is used.
Souffl{\'e} rejects a program that violates this condition. Reachability is the standard
example. Over an input relation $\textit{Edge}$, the program
\[
\textit{Reachable}(X, Y) \horn \textit{Edge}(X, Y). \qquad
\textit{Reachable}(X, Z) \horn \textit{Edge}(X, Y), \textit{Reachable}(Y, Z).
\]
computes the transitive closure as the least fixed point of the two rules.
No fixed number of joins over $\textit{Edge}$ can express it, since paths may be arbitrarily long.

\noindent\textbf{Demands specific to Datalog synthesis.}
Three properties make writing such a program from a natural-language question demanding. 
\emph{Recursive fixed-point reasoning}: a recursive program is correct only if its base case,
its inductive step, and the variable bindings between them together yield the intended
least fixed point. \emph{Predicate invention}: a solution may introduce auxiliary
predicates absent from the schema to express a decomposition; in Souffl{\'e},
it must name and declare each such predicate with typed attributes. \emph{Static checking}: Souffl{\'e} checks the whole program
before evaluating it, so one mis-bound variable, one type clash between joined relations, or one
stratification violation rejects a program that would otherwise run. Such a program never reaches
the data, so compile failures and semantic errors are distinct outcomes.

\begin{figure}[t]
	\centering
	\includegraphics[width=0.95\linewidth]{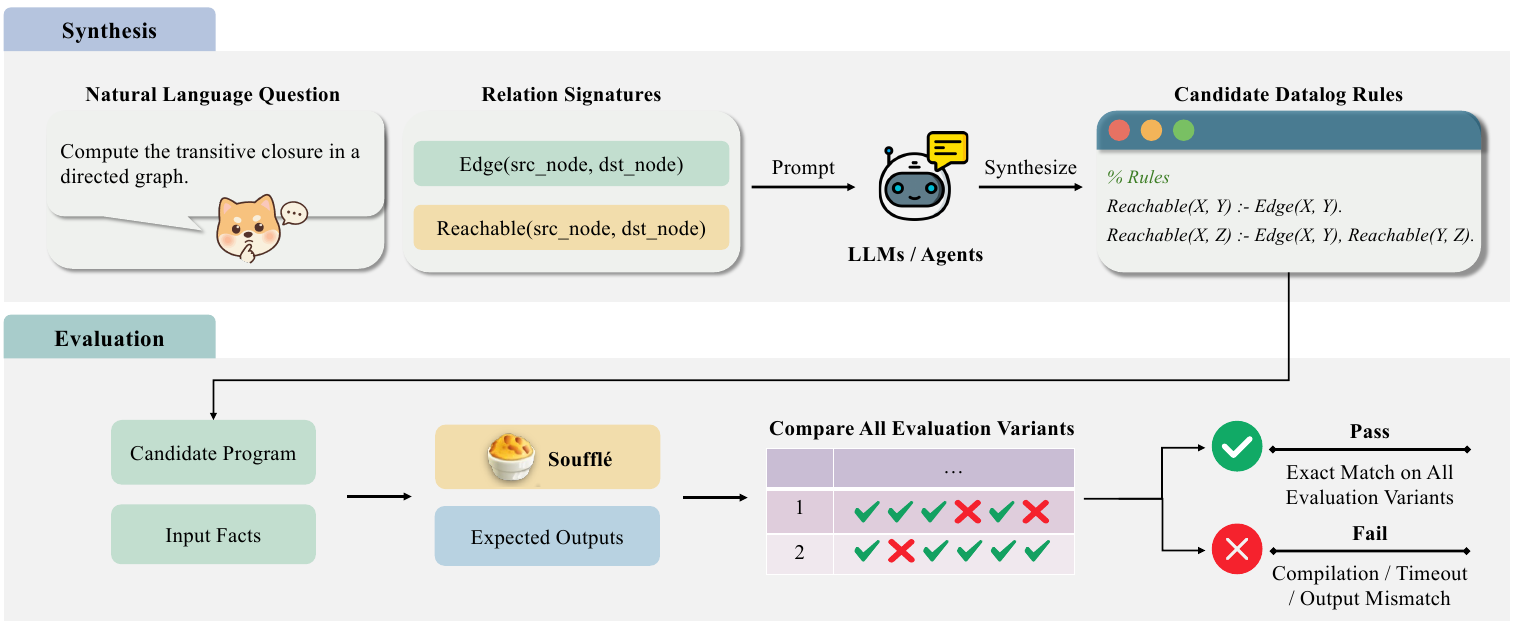}
	\caption{A \bench task with relation signatures only. The prompt carries the question
	and the signatures, and Souffl{\'e} runs the candidate program on the held-out evaluation
	variants and compares its output with the expected facts.}
	\label{fig:task}
\end{figure}

\noindent\textbf{Task formulation.}
\label{sec:task}
Each task in \bench is an end-to-end text-to-Datalog synthesis
problem, a tuple
\begin{equation}
	\tau_i = \left(q_i,\ \mathcal{R}_i^{\mathrm{in}},\ \mathcal{R}_i^{\mathrm{out}},\ d_i,\ \mathcal{D}_i,\ \mathcal{T}_i\right),
	\label{eq:task}
\end{equation}
where $q_i$ is the natural-language question, $\mathcal{R}_i^{\mathrm{in}}$ and
$\mathcal{R}_i^{\mathrm{out}}$ are the input and output relation signatures, and
$d_i$ is an optional set of relation descriptions. $\mathcal{D}_i$ is the
\emph{demonstration instance}, which may appear in a prompt as an input--output example or
supply execution feedback to an iterative system, and $\mathcal{T}_i$ is a disjoint
set of \emph{evaluation variants} used only to compute scores. Each instance pairs
EDB facts $x_{ij}$ with the expected IDB output $y_{ij}$. Given a prompt carrying
the schema $\left(\mathcal{R}_i^{\mathrm{in}},\ \mathcal{R}_i^{\mathrm{out}}\right)$ and,
optionally, $d_i$, a system produces a candidate program $\hat{P}_i$, which may introduce
auxiliary predicates absent from the schema. An iterative system may adapt to
$\mathcal{D}_i$ but never to $\mathcal{T}_i$, and every reported metric is
computed on $\mathcal{T}_i$ after the candidate is frozen. Figure~\ref{fig:task} follows
the reachability task above through both stages: synthesis from the question and relation
signatures, and grading of the candidate on every evaluation variant. Appendix~\ref{sec:pre}
gives the formal synthesis problem this instantiates.

\section{\bench}

This section describes how the tasks were collected and specified,
characterizes the resulting benchmark, and validates its execution oracle.

\subsection{Benchmark Construction}
\label{sec:construction-phases}

\begin{figure}[b]
	\centering
	\includegraphics[width=0.9\linewidth]{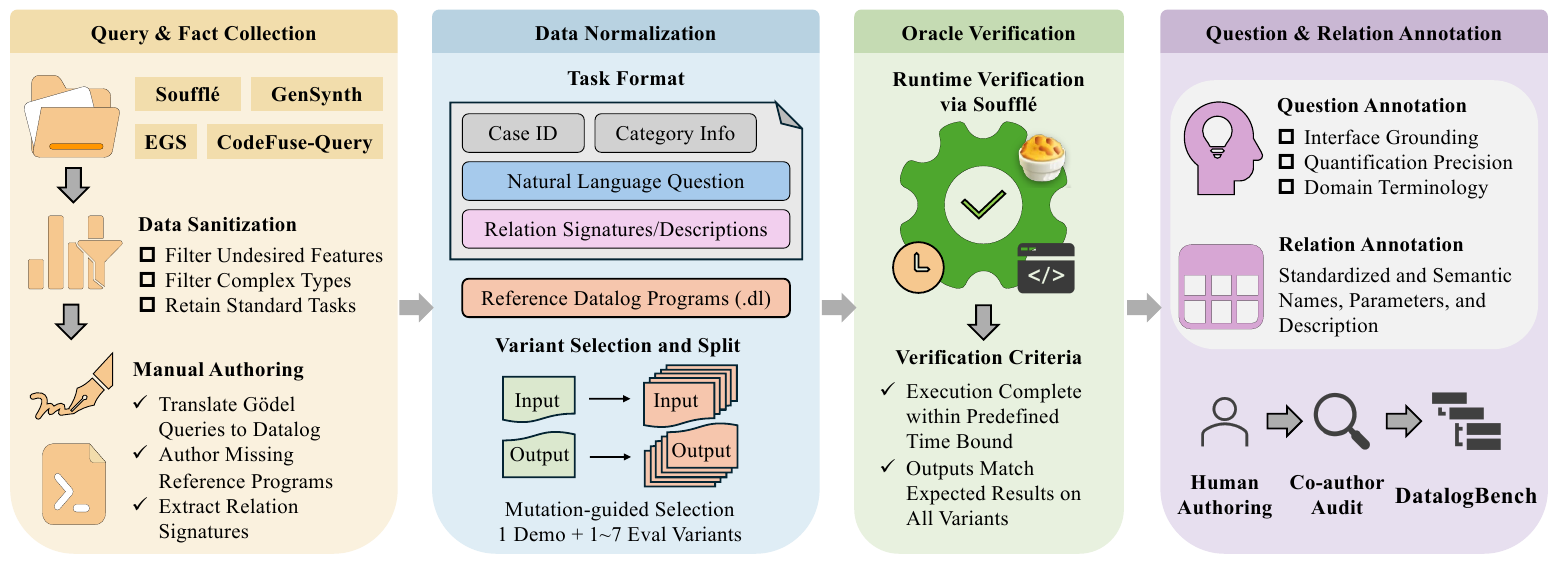}
	\caption{The four phases of benchmark construction and what each produces.}
	\label{fig:bench_con}
	\vspace{-3mm}
\end{figure}

\bench was assembled in four phases, each securing one property of the tasks (Figure~\ref{fig:bench_con}). \emph{Collection} draws reference programs and their input--output
facts from four public artifacts: Souffl{\'e}~\citep{jordan2016souffle}, EGS~\citep{thakkar2021example}, GenSynth~\citep{mendelson2021gensynth}, and
CodeFuse-Query~\citep{xie2024codefuse}, whose queries we rewrote in Souffl{\'e}, so that every task originates in a program written for an actual analysis or test.
Because every source is public, Appendix~\ref{app:contamination} measures what exposure to it is worth.
\emph{Normalization} renames relations, standardizes argument types, and regenerates input facts.
As the upstream names no longer appear, a model must work from the schema
in the prompt. \emph{Oracle verification} re-executes every reference program under
Souffl{\'e} and keeps a task only when its outputs match the recorded expected outputs
exactly. \emph{Annotation} adds a human-authored question and relation
descriptions, requiring that every output relation be named in the question and
that quantifiers be explicit (Appendix~\ref{app:construction}).

The specification deliberately omits auxiliary predicates, leaving predicate invention to the model. A task may admit many correct programs, but its question
and schema must together fix a single intended output. During construction,
the authors audited the questions and relation descriptions against the reference
programs and revised the annotations. Twenty-five tasks are determinate only given
domain vocabulary that a schema cannot carry. Following BIRD~\citep{li2023can}, for the fifteen whose vocabulary we judged
beyond common knowledge, a knowledge note supplies it without hinting at program structure
(Appendix~\ref{app:setup-details}); the other ten assume a reader who already has it.

\subsection{Benchmark Statistics}
\label{sec:stats}

\begin{table}[t]
	\centering
	\small
	\caption{Structural and data statistics of \bench.
	The largest arity of a program is the most arguments any of its predicates takes; strata are the layers of a stratified evaluation;
	an SCC is a set of mutually recursive predicates, and its count is the most recursive SCCs in any one program;
	trimmed averages exclude the five largest tasks. Tuple counts are for each task's first evaluation variant.}
	\label{tab:complex}
	\begin{tabular}{lc@{\hskip 2.5em}lc}
		\toprule
		\multicolumn{2}{c}{\textbf{Program complexity}} & \multicolumn{2}{c}{\textbf{Datalog-specific structure}} \\
		\cmidrule(lr){1-2}\cmidrule(lr){3-4}
		Avg. / max rules       & 5.1 / 57  & Recursive tasks                & 78 \\
		Avg. / max body atoms  & 2.3 / 10  & Max recursive SCC size / count & 9 / 4 \\
		Avg. / max largest arity & 3.4 / 20 & Avg. auxiliary predicates      & 1.4 \\
		Avg. / max strata      & 1.5 / 5   & Tasks using negation / aggregation & 34 / 22 \\
		\midrule
		\multicolumn{2}{c}{\textbf{Inputs}} & \multicolumn{2}{c}{\textbf{Outputs}} \\
		\cmidrule(lr){1-2}\cmidrule(lr){3-4}
		Avg. relations          & 2.9       & Avg. relations                 & 1.5 \\
		Median tuples           & 10        & Median tuples                  & 5 \\
		Trimmed avg. tuples     & 12.5      & Trimmed avg. tuples            & 12.3 \\
		\midrule
		\multicolumn{4}{c}{\textbf{Evaluation variants}} \\
		\cmidrule(lr){1-4}
		Total                   & 251       & Range per task                 & 1--7 \\
		Mean / median per task  & 1.85 / 2  & Single-variant tasks           & 57 \\
		\bottomrule
	\end{tabular}
\end{table}
Table~\ref{tab:complex} summarizes the structure of the reference programs.
Two properties matter most for what follows. First, $78$ of the $136$ reference programs are recursive and $59$ introduce at least one auxiliary predicate.
Second, all $136$ reference programs are stratified, since Souffl{\'e} compiles every one of them,
so any stratification error in a candidate is the candidate's own. Data scale is heavy-tailed (means of $82$ input and
$1{,}517$ output tuples, against medians of $10$ and $5$). We keep the tail because tasks such as pointer analysis and
large-graph reachability show their characteristic behavior only at non-trivial sizes.
The four domains---program analysis ($n=47$), graph analytics ($n=33$), formal
reasoning ($n=29$), and knowledge discovery ($n=27$)---reflect the application area of each task, and they
differ in structure: recursion appears in $79.3\%$ of formal reasoning and $75.8\%$ of graph analytics tasks but in
only $42.6\%$ of program analysis and $37.0\%$ of knowledge discovery tasks,
and knowledge discovery introduces $0.6$ auxiliary predicates per task
against $1.6$ elsewhere. We also report evaluation results
by structural property (\S\ref{sec:failure}). Appendix~\ref{app:stats} gives these
statistics in full, broken down by domain, and further scores task difficulty along
two structural axes.

\subsection{Benchmark Validation}
\label{sec:validation}

\noindent\textbf{Held-out variants guard against overfitting.}
Every task separates the demonstration instance, which may appear in a prompt as an input--output example
or supply execution feedback to an iterative system, from the evaluation variants used for scoring.
A program fitted to the demonstration instance therefore earns credit only if it also computes
the intended relations on inputs it has not seen; \S\ref{sec:agent} shows that both coding
agents produce candidates that fail this test.

\noindent\textbf{Mutation analysis measures oracle discrimination.}
We test the evaluation variants against mutants of the reference programs
generated by rule deletion, body-atom deletion, and join-variable rebinding.
They distinguish $1{,}440$ of $1{,}521$ compilable mutants ($94.7\%$), compared
with $1{,}246$ using only the first variant of each task.
To cover additional error types, we apply five further operators,
for which the variants distinguish $693$ of $800$ compilable mutants
($86.6\%$).
Equivalent and unresolved mutants remain in the denominators; these rates
measure discrimination within the tested mutation families.
Appendix~\ref{app:audit} provides operator-level results, survivor analysis,
and additional-input checks.

\section{Evaluation}
\label{sec:eval}

\subsection{Experimental Setup}
\label{sec:setup}

\noindent\textbf{Models.}
We evaluate six LLMs from four providers (Table~\ref{tab:model-inventory}):
GPT~5.6~Sol~\citep{openai2026gpt56}, Claude Opus~5~\citep{anthropic2026opus5}, Gemini~3.8~Flash~\citep{google2026gemini38flash},
DeepSeek~V4~Pro, and DeepSeek~V4~Flash~\citep{deepseek2026v4}, the latter in \emph{thinking} and
\emph{non-thinking} modes, written DeepSeek~V4~FT and DeepSeek~V4~FNT below,
which isolate test-time reasoning on one underlying model.
All share one untuned prompt template (Appendix~\ref{app:prompts}). Appendix~\ref{app:model-details} records
the request contracts. Candidate programs
run under Souffl{\'e}~2.5 with a $30$-second timeout (Appendix~\ref{app:artifact}).

\noindent\textbf{Direct prompting (\S\ref{sec:prompt-enrichment}).}
We prompt each model directly with one completion per task in each run, varying
the prompt along two axes: the schema is given as relation signatures alone
(names and argument types) or together with natural-language descriptions of each
relation, and the demonstration instance is either withheld or appended as an
input--output example. We call the resulting $24$ model--prompt combinations
the \emph{Direct} cells. Every cell is scored on the same held-out
evaluation variants (\S\ref{sec:task}), and the fifteen tasks with a knowledge note
carry it in every prompt.

\noindent\textbf{Coding agents (\S\ref{sec:agent}).}
We evaluate Codex CLI~\citep[Codex;][]{openai2025codexcli} with GPT~5.6~Sol
and Claude Code~\citep[CC;][]{anthropic2025claudecode} with Claude
Opus~5 on relation signatures without an input--output example, the same prompt as the
Direct cells they are compared with.
Main agent results use the first four turns of a persistent session. Agents may
use local file and shell tools, but not web search or retrieval. After each turn, the harness runs
the candidate on the demonstration instance and returns either the compile error
or a bounded sample of differing tuples. The task stops once the candidate is
exact on the demonstration instance. Evaluation variants stay held out until the candidate is frozen.
Each agent is compared with the Direct cell of its own model, which holds the model fixed,
and we do not compare the two agents, whose models and clients both differ.
Appendix~\ref{app:model-details} gives the enforcement and environment details.
Resource use is reported in Appendix~\ref{app:agent-detail}.

\noindent\textbf{Statistical inference.}
Each configuration has \nruns{} reported runs. We summarize per-run scores by
their mean and, where shown, sample standard deviation, which describes
variability across runs. For inference, we first average outcomes across runs
within each task and use tasks as the resampling units. We compute $95\%$
percentile-bootstrap confidence intervals using $10{,}000$ resamples with fixed
random seeds. For paired comparisons between configurations, we compute
differences within each task and apply a two-sided sign-flip test to these
task-level differences.

\subsection{Evaluation Metrics}
\label{sec:compute}
A candidate program is graded by execution. For task $i$ in the evaluated set
$\mathcal{T}$, let $C_i=1$ if the candidate compiles and $0$ otherwise,
and let $M_i=1$ if it also reproduces the expected output on all $m_i$ of that
task's evaluation variants. \emph{Compile Pass} (CP) and \emph{Exact Match} (EX)
are the task averages of the two:
\begin{equation}
	\mathrm{CP}(\mathcal{T}) = \frac{1}{|\mathcal{T}|}\sum_{i \in \mathcal{T}} C_i,
	\qquad
	\mathrm{EX}(\mathcal{T}) = \frac{1}{|\mathcal{T}|}\sum_{i \in \mathcal{T}} M_i.
	\label{eq:cpex}
\end{equation}
CP measures whether a program passes Souffl{\'e}'s static checks, and EX measures
whether it is semantically correct: a task counts only if its program runs
\emph{and} agrees with the reference on every held-out variant. Appendix~\ref{app:metrics}
gives the full form and the secondary metrics.

Souffl{\'e} requires an explicit typed \texttt{.decl} for every relation,
including auxiliary predicates, so a program that derives a relation by rule but
never declares it is rejected before it runs. This requirement is a property of
the dialect rather than of the task. To measure its contribution without
conflating it with rule repair, we also report \emph{repaired exact match}
(EX$^\dagger$). A task counts if its program is exact as written, or becomes
exact once declarations omitted for auxiliary predicates are synthesized from the
program's own rules: argument types are taken from relations that bind each
variable, and no rule is added, removed, or altered. EX$^\dagger$ serves only as a diagnostic;
raw EX remains the primary measure of what the model submitted.

\section{Results}
\label{sec:result}

\subsection{Direct Prompting}
\label{sec:prompt-enrichment}

\begin{table}[t]
	\centering
	\caption{Zero-shot results with relation signatures or descriptions, as mean $\pm$
	standard deviation over \nruns{} runs.
	EX$^\dagger$ also credits programs made exact by synthesizing only
	their omitted declarations. Bold marks each column's best.}
	\label{tbl:base}
		\resizebox{\textwidth}{!}{%
	\begin{tabular}{lccc|ccc}
			\toprule
			\multirow{2}{*}{Model} & \multicolumn{3}{c}{Signature} &
			\multicolumn{3}{c}{Description} \\
			\cmidrule(lr){2-4}\cmidrule(lr){5-7}
			& CP (\%) & EX (\%) & EX$^\dagger$ (\%) & CP (\%) & EX (\%) & EX$^\dagger$ (\%) \\
			\midrule
			GPT~5.6~Sol             & $74.3 \pm 0.7$ & $53.2 \pm 1.5$ & 59.8 & $75.5 \pm 1.7$ & $55.4 \pm 2.1$ & 61.3 \\
			Claude Opus~5           & $85.5 \pm 1.1$ & $64.2 \pm 0.4$ & 64.2 & $83.8 \pm 2.7$ & $63.0 \pm 2.4$ & 63.0 \\
			Gemini~3.8~Flash        & $\mathbf{87.5} \pm 0.0$ & $\mathbf{65.7} \pm 1.1$ & \textbf{69.9} & $\mathbf{86.3} \pm 0.4$ & $\mathbf{67.2} \pm 2.4$ & \textbf{69.9} \\
			DeepSeek~V4~Pro         & $62.5 \pm 2.2$ & $45.6 \pm 2.5$ & 54.7 & $62.0 \pm 3.3$ & $49.0 \pm 2.8$ & 57.1 \\
			DeepSeek~V4~FT          & $60.0 \pm 1.1$ & $45.3 \pm 2.6$ & 57.1 & $60.8 \pm 3.3$ & $48.3 \pm 2.1$ & 58.8 \\
			DeepSeek~V4~FNT         & $67.9 \pm 2.2$ & $43.4 \pm 1.5$ & 43.4 & $71.1 \pm 2.6$ & $46.3 \pm 1.9$ & 46.3 \\
			\bottomrule
		\end{tabular}}
\end{table}

\noindent\textbf{Direct prompting remains unreliable.}
Across the six models, zero-shot EX averages $52.9\%$ with signatures and
$54.9\%$ with descriptions (Table~\ref{tbl:base}; Appendix~\ref{app:per-domain} splits it by domain). CP is higher ($73.0\%$ and $73.2\%$), so about
a fifth of all programs compile yet compute the wrong relations (\S\ref{sec:failure}).
Performance also varies substantially by model: Gemini~3.8~Flash reaches
$67.2\%$ with descriptions, while DeepSeek~V4~FNT reaches only $43.4\%$ with signatures.
Across all $24$ Direct cells, the best results are Claude~Opus~5 with one
input--output example, at $75.7\%$ EX under either schema.

\begin{figure}[t]
	\centering

	\includegraphics[width=0.88\linewidth]{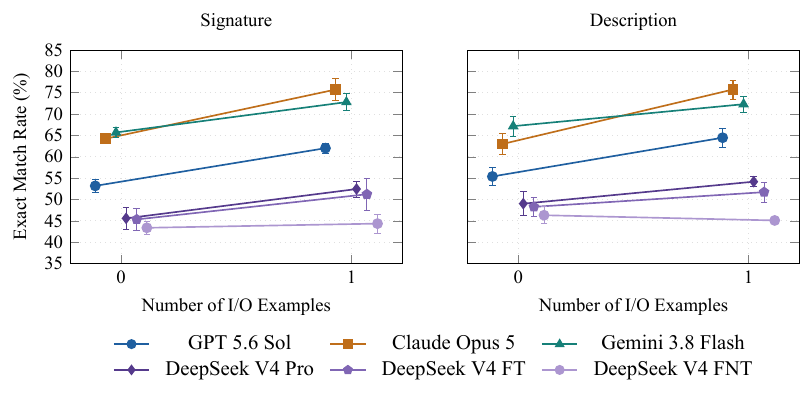}
	\caption{Exact match with and without one input--output example, under each schema.
	Points are means over \nruns{} runs and error bars show one standard deviation
	across runs; the models are offset slightly along the horizontal axis.}
	\label{fig:fewshot}
	\vspace{-2mm}
\end{figure}

\noindent\textbf{An example helps on average; descriptions add little.}
Across models, adding one input--output example increases mean EX by $6.9$
percentage points with signatures and $5.7$ with descriptions ($95\%$ CI
$[3.1,10.6]$ and $[2.3,9.1]$, respectively; Figure~\ref{fig:fewshot}). Replacing
signatures with descriptions yields a smaller mean gain of $1.4$ points,
averaged over models and example conditions ($[-0.1,3.0]$, paired sign-flip
$p=.0879$). At the individual-model level, $95\%$ intervals exclude zero for
eight of the twelve example contrasts, all positive, and for none of the twelve
description-versus-signature contrasts.

For reference, symbolic synthesizers that learn from input--output examples reach
at most $36.0\%$ EX under a protocol that favors them (Appendix~\ref{app:symbolic}).

\subsection{Coding Agents}
\label{sec:agent}

\begin{table}[t]
	\centering
	\caption{Coding agents with relation signatures, no input--output example, and the first four turns (\%), as means over \nruns{} runs; Overall also gives the sample standard deviation.
	KD: knowledge discovery; GA: graph analytics; PA: program analysis; FR: formal reasoning.}
	\label{tab:agent}
	\resizebox{0.98\textwidth}{!}{
		\begin{tabular}{lcccccccccc}
			\toprule
			\multirow{2}{*}{Agent / Model} &
			\multicolumn{5}{c}{Compile Pass} &
			\multicolumn{5}{c}{Exact Match} \\
			\cmidrule(lr){2-6}\cmidrule(lr){7-11}
			& KD & GA & PA & FR & \textbf{Overall} & KD & GA & PA & FR & \textbf{Overall} \\
			\midrule
			Codex / GPT~5.6~Sol & 100.0 & 99.0 & 100.0 & 100.0 & $99.8\pm0.4$ & 88.9 & 81.8 & 80.1 & 83.9 & $83.1\pm1.3$ \\
			CC / Claude Opus~5 & 100.0 & 100.0 & 99.3 & 97.7 & $99.3\pm1.3$ & 90.1 & 89.9 & 85.8 & 81.6 & $86.8\pm2.2$ \\
			\bottomrule
		\end{tabular}
	}
\end{table}

\noindent\textbf{Agents substantially outperform their matched Direct cells.}
Codex improves over GPT~5.6~Sol from $53.2\%$ to $83.1\%$ EX (Table~\ref{tab:agent}), a paired gain
of $29.9$ points (bootstrap $95\%$ CI $[23.0,37.0]$). CC improves over
Claude~Opus~5 from $64.2\%$ to $86.8\%$, a gain of $22.5$ points
($[16.4,28.9]$). Both paired sign-flip tests give $p<.001$, and both gains
appear in every run.
The agents receive execution feedback on the demonstration instance in a
persistent session with file and shell tools.
\begin{figure}[t]
	\centering

	\includegraphics[width=0.92\linewidth]{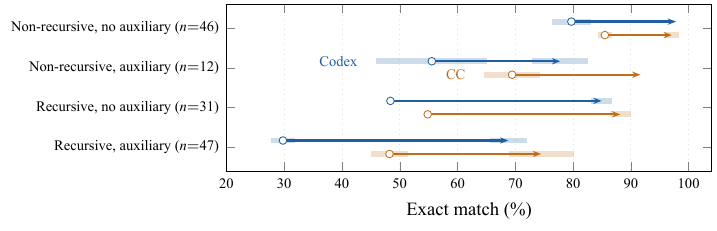}
	\caption{Exact match by reference structure, from each model's Direct cell
	(circle) to its agent (arrow head). \emph{Auxiliary} means the reference
	introduces at least one predicate absent from the schema.
	Positions are means over \nruns{} runs; shaded bands span one standard
	deviation across runs.}
	\label{fig:structure}
	\vspace{-2mm}
\end{figure}

\noindent\textbf{Interaction narrows but does not remove structural gaps.}
Among tasks solved by the agent but not its matched Direct cell, Codex gains
$41.0$ per run and CC gains $32.0$. Declaration repair alone accounts for $9.0$ of Codex's gains
and none of CC's; the remaining gains are concentrated in recursive tasks
($23.7/32.0$ for both agents) and tasks whose reference programs use auxiliary predicates
($14.3/32.0$ and $16.3/32.0$). Figure~\ref{fig:structure} shows the $47$ tasks whose references have
both properties rising to $68.8\%$ for Codex and $74.5\%$ for CC, still well below the $97.8\%$ and $97.1\%$
they reach on the $46$ tasks whose references have neither property. The agents therefore make
progress on the structural challenges without saturating the benchmark.

Feedback on one demonstration instance can also overfit it. Of the candidates
each agent returned as exact on the demonstration instance ($129.0$ for Codex and
$131.7$ for CC per run), the held-out variants reject $16.7$ and $15.0$. Separately,
two CC sessions in one run time out after the retry budget fixed in advance and
count as zero (Appendix~\ref{app:agent-detail}).

\subsection{Failure Mode Analysis}
\label{sec:failure}

\noindent\textbf{Most failures under direct prompting occur at compile time.}
Across the $9{,}792$ Direct evaluations in the reported runs, $27.2\%$
fail to compile and $14.8\%$ compile but compute the wrong relations;
$0.8\%$ time out during evaluation and $0.2\%$ terminate without a program.
Across the six models' best cells, compile failures account for
$13.5\%$--$39.0\%$ of tasks and compiled-but-wrong programs for $7.6\%$--$23.8\%$. We classify each compile failure by the statement the
compiler rejects (Appendix~\ref{app:error-taxonomy}). The largest group, $53.7\%$,
is predicate invention left incomplete: a rule uses an auxiliary predicate the
program never declares ($986$), or the compiler cannot type a rule that defines
one ($445$). Souffl{\'e} dialect errors follow at $25.8\%$, such as aggregates in
rule heads and negation or aggregate syntax from other languages, then the typing
of constants and schema relations ($12.8\%$) and variable safety and
stratification ($5.8\%$). Appendix~\ref{app:qualitative} shows representative failing programs.
Of $885$ Direct programs ($9.0\%$) whose compile errors concern only undefined
auxiliary predicates, inferring declarations without changing any rule makes
$856$ compile and $534$ exact. This raises mean EX across the $24$ Direct cells
by $5.5$ points (the EX$^\dagger$ column of Table~\ref{tbl:base} shows zero-shot results).

\noindent\textbf{Recursion and decomposition remain the main difficulty.}
With relation signatures and no example, EX ranges from $23.5\%$ to $51.3\%$
on the $78$ recursive tasks, versus $67.8\%$--$85.1\%$ on the $58$ non-recursive
tasks. On the $59$ tasks whose references use auxiliary predicates, EX is
$20.9\%$--$53.1\%$, versus $60.2\%$--$75.3\%$ on the other $77$.
Each property separates outcomes with the other held fixed. The $47$ tasks whose references have both properties reach $17.7\%$--$48.9\%$ EX,
compared with $73.9\%$--$89.1\%$ for the $46$ with neither.
After adjustment for program complexity (Appendix~\ref{sec:analysis}),
the Direct-cell gaps remain at $-27.8$ points for recursion and $-18.1$ for
auxiliary predicates.

The task groups also differ in failure type. Auxiliary predicates mainly raise
compile failures: from $10.4\%$ to $29.2\%$ on non-recursive tasks and from
$17.4\%$ to $49.6\%$ on recursive tasks. Recursion adds semantic errors:
recursive tasks without auxiliary predicates are the only group with more
compiled-but-wrong programs than compile failures ($25.8\%$ versus $17.4\%$).

\noindent\textbf{The agents remove compile failures but fewer semantic errors.}
Against their matched Direct cells, the agents cut compile failures per run from $35.0$ to $0.3$ (Codex)
and from $19.7$ to $0.3$ (CC), but programs that compile yet are not exact only from $28.7$ to $22.7$
and from $29.0$ to $17.0$. What remains is therefore mostly semantic, and it concentrates in
recursive tasks: $19.0$ of Codex's $22.7$ and $14.7$ of CC's $17.0$.

These findings suggest a division of labor (Appendix~\ref{sec:discuss}). Most compile failures are mechanical:
deterministic checks or a compiler in the loop can remove them, as the agents
do, and declaration inference alone recovers part of the loss. The failures
that survive are semantic and sit in recursion, so model- or search-based repair
should focus on base cases, inductive bindings, and constructing useful
decompositions through auxiliary predicates. Execution feedback is valuable, but correctness on a single
demonstration instance is not a substitute for held-out evaluation.

\section{Conclusion}
\label{sec:conclusion}
We introduced \bench, a benchmark of text-to-Datalog synthesis tasks in the
Souffl{\'e} dialect, together with an execution oracle validated by mutation
analysis. Across six LLMs and two coding agents,
direct prompting remains unreliable; an input--output example helps most models,
while relation descriptions add little. Most failures of direct prompting occur at compile
time, typically involving auxiliary predicates that the model invents but never declares
or types consistently. Supplying missing declarations recovers part
of exact match. The coding agents eliminate nearly all compile failures and raise
exact match substantially, but the errors they leave are mostly semantic, concentrated in
recursive tasks, and the tasks that combine recursion with auxiliary predicates
remain clearly harder. The open challenges are therefore recursive reasoning and
decomposition: constructing the right fixed point and inventing auxiliary
predicates for useful decompositions.

\textbf{Limitations.} This study has three main limitations.
First, \bench targets the Souffl{\'e} dialect alone, and we do not measure how far the findings transfer.
Second, the execution oracle is bounded by mutation analysis rather than proved complete.
Third, the agent setting supplies execution feedback and interactive tooling together, so their contributions are not separated.
Appendix~\ref{app:limitations} discusses these limitations in full.

\section*{AI Use Statement}

LLMs are the object of study in this work: they are evaluated as Datalog
synthesizers, and their generations are the data we analyze. Separately from
that role, no part of the released benchmark is machine-generated. All
natural-language questions and relation descriptions were written by the
authors, and the reference programs were either taken from the upstream
artifacts listed in \S\ref{sec:construction-phases} or written by the
authors, so the specifications the models are scored against are not drawn from
the distribution of the models being scored. We used LLMs to assist with grammar
editing and language polishing, and LLM-based coding agents to write analysis and
plotting scripts from measurements we supplied. No benchmark content was
generated by an LLM, and every number reported in this paper comes from running
the released evaluation pipeline.

\section*{Ethics Statement}

\bench is assembled from publicly released research artifacts and open-source
tooling: Souffl{\'e} (UPL-1.0), EGS (MIT), and CodeFuse-Query (Apache-2.0), each
used under its stated license, and GenSynth, whose repository declares no
license; we reference GenSynth as an unmodified submodule and derive task
specifications from its published benchmark files. The benchmark contains only
synthetic relational facts and program-analysis fixtures; it includes no personal
data, no user-generated content, and no material about identifiable individuals.
All questions, relation descriptions, and annotations were written by the
authors, with no crowdworkers involved. The evaluated systems are commercial LLM
APIs and coding agents queried through their public interfaces. Generated Datalog
is executed by Souffl{\'e} in a scratch directory, and the coding agents run in
per-task scratch directories inside a pinned container rather than in the
repository. The benchmark tree and reference programs are not mounted, outbound
traffic is forced through a proxy that allows only the model endpoint, and each
client's web tools are disabled (\S\ref{sec:setup}; Appendix~\ref{app:contracts}). No generated program
writes outside its scratch directory. We see no dual-use concern specific to this
work: the capability under study is writing declarative database queries.

\section*{Reproducibility Statement}

The benchmark, prompt templates, evaluation harness, and scoring and analysis
scripts are available at \repourl.
Section~\ref{sec:setup} states
the decoding configuration, the Souffl{\'e} version, and the per-execution
timeout, and its paragraph on statistical inference specifies the bootstrap and
permutation procedures, including the resample count and the seeding. Section~\ref{sec:compute} defines the metrics.
Appendix~\ref{app:prompts} reproduces the prompt templates, and
Appendix~\ref{app:model-details} lists every evaluated model, agent, and symbolic
baseline together with the configuration each was run under. The symbolic
baselines are included as pinned submodules with the source-level build patches
required on a modern toolchain, and are built by a single setup script that needs
no container runtime. Structural statistics of the reference programs are
produced by an analyzer in the repository, so the numbers in
\S\ref{sec:stats} can be regenerated from the benchmark files alone, and
every derived figure in the main text and appendix---the declaration repair, the
decomposition of the agents' gains, the contamination scan, and the overlap with
symbolic synthesis---is produced by a script in the repository.

\bibliographystyle{preprint}
\bibliography{llm}

@book{abiteboul1995foundations,
	author    = {Serge Abiteboul and
	Richard Hull and
	Victor Vianu},
	title     = {Foundations of Databases},
	publisher = {Addison-Wesley},
	year      = {1995},
	isbn      = {0-201-53771-0},
	bibsource = {DBLP, http://dblp.uni-trier.de}
}

@inproceedings{bravenboer2009strictly,
	author       = {Martin Bravenboer and
	Yannis Smaragdakis},
	editor       = {Shail Arora and
	Gary T. Leavens},
	title        = {Strictly declarative specification of sophisticated points-to analyses},
	booktitle    = {Proceedings of the 24th Annual {ACM} {SIGPLAN} Conference on Object-Oriented
	Programming, Systems, Languages, and Applications, {OOPSLA} 2009,
	October 25-29, 2009, Orlando, Florida, {USA}},
	pages        = {243--262},
	publisher    = {{ACM}},
	year         = {2009},
	url          = {https://doi.org/10.1145/1640089.1640108},
	doi          = {10.1145/1640089.1640108},
	bibsource    = {dblp computer science bibliography, https://dblp.org}
}

@inproceedings{whaley2004cloning,
	author       = {John Whaley and
	Monica S. Lam},
	editor       = {William W. Pugh and
	Craig Chambers},
	title        = {Cloning-based context-sensitive pointer alias analysis using binary
	decision diagrams},
	booktitle    = {Proceedings of the {ACM} {SIGPLAN} 2004 Conference on Programming
	Language Design and Implementation 2004, Washington, DC, USA, June
	9-11, 2004},
	pages        = {131--144},
	publisher    = {{ACM}},
	year         = {2004},
	url          = {https://doi.org/10.1145/996841.996859},
	doi          = {10.1145/996841.996859},
	bibsource    = {dblp computer science bibliography, https://dblp.org}
}

@inproceedings{loo2006declarative,
	author       = {Boon Thau Loo and
	Tyson Condie and
	Minos N. Garofalakis and
	David E. Gay and
	Joseph M. Hellerstein and
	Petros Maniatis and
	Raghu Ramakrishnan and
	Timothy Roscoe and
	Ion Stoica},
	editor       = {Surajit Chaudhuri and
	Vagelis Hristidis and
	Neoklis Polyzotis},
	title        = {Declarative networking: language, execution and optimization},
	booktitle    = {Proceedings of the {ACM} {SIGMOD} International Conference on Management
	of Data, Chicago, Illinois, USA, June 27-29, 2006},
	pages        = {97--108},
	publisher    = {{ACM}},
	year         = {2006},
	url          = {https://doi.org/10.1145/1142473.1142485},
	doi          = {10.1145/1142473.1142485},
	bibsource    = {dblp computer science bibliography, https://dblp.org}
}

@inproceedings{alvaro2010boom,
	author       = {Peter Alvaro and
	Tyson Condie and
	Neil Conway and
	Khaled Elmeleegy and
	Joseph M. Hellerstein and
	Russell Sears},
	editor       = {Christine Morin and
	Gilles Muller},
	title        = {{BOOM} analytics: exploring data-centric, declarative programming for
	the cloud},
	booktitle    = {European Conference on Computer Systems, Proceedings of the 5th European
	conference on Computer systems, EuroSys 2010, Paris, France, April
	13-16, 2010},
	pages        = {223--236},
	publisher    = {{ACM}},
	year         = {2010},
	url          = {https://doi.org/10.1145/1755913.1755937},
	doi          = {10.1145/1755913.1755937},
	bibsource    = {dblp computer science bibliography, https://dblp.org}
}

@inproceedings{halperin2014demonstration,
	author       = {Daniel Halperin and
	Victor Teixeira de Almeida and
	Lee Lee Choo and
	Shumo Chu and
	Paraschos Koutris and
	Dominik Moritz and
	Jennifer Ortiz and
	Vaspol Ruamviboonsuk and
	Jingjing Wang and
	Andrew Whitaker and
	Shengliang Xu and
	Magdalena Balazinska and
	Bill Howe and
	Dan Suciu},
	editor       = {Curtis E. Dyreson and
	Feifei Li and
	M. Tamer {\"{O}}zsu},
	title        = {Demonstration of the {Myria} big data management service},
	booktitle    = {International Conference on Management of Data, {SIGMOD} 2014, Snowbird,
	UT, USA, June 22-27, 2014},
	pages        = {881--884},
	publisher    = {{ACM}},
	year         = {2014},
	url          = {https://doi.org/10.1145/2588555.2594530},
	doi          = {10.1145/2588555.2594530},
	bibsource    = {dblp computer science bibliography, https://dblp.org}
}

@inproceedings{seo2013socialite,
	author       = {Jiwon Seo and
	Stephen Guo and
	Monica S. Lam},
	editor       = {Christian S. Jensen and
	Christopher M. Jermaine and
	Xiaofang Zhou},
	title        = {{SociaLite}: {Datalog} extensions for efficient social network analysis},
	booktitle    = {29th {IEEE} International Conference on Data Engineering, {ICDE} 2013,
	Brisbane, Australia, April 8-12, 2013},
	pages        = {278--289},
	publisher    = {{IEEE} Computer Society},
	year         = {2013},
	url          = {https://doi.org/10.1109/ICDE.2013.6544832},
	doi          = {10.1109/ICDE.2013.6544832},
	bibsource    = {dblp computer science bibliography, https://dblp.org}
}

@inproceedings{shkapsky2016big,
	author       = {Alexander Shkapsky and
	Mohan Yang and
	Matteo Interlandi and
	Hsuan Chiu and
	Tyson Condie and
	Carlo Zaniolo},
	editor       = {Fatma {\"{O}}zcan and
	Georgia Koutrika and
	Sam Madden},
	title        = {Big Data Analytics with {Datalog} Queries on {Spark}},
	booktitle    = {Proceedings of the 2016 International Conference on Management of
	Data, {SIGMOD} Conference 2016, San Francisco, CA, USA, June 26 -
	July 01, 2016},
	pages        = {1135--1149},
	publisher    = {{ACM}},
	year         = {2016},
	url          = {https://doi.org/10.1145/2882903.2915229},
	doi          = {10.1145/2882903.2915229},
	bibsource    = {dblp computer science bibliography, https://dblp.org}
}

@inproceedings{albarghouthi2017constraint,
	author       = {Aws Albarghouthi and
	Paraschos Koutris and
	Mayur Naik and
	Calvin Smith},
	editor       = {J. Christopher Beck},
	title        = {Constraint-Based Synthesis of {Datalog} Programs},
	booktitle    = {Principles and Practice of Constraint Programming - 23rd International
	Conference, {CP} 2017, Melbourne, VIC, Australia, August 28 - September
	1, 2017, Proceedings},
	series       = {Lecture Notes in Computer Science},
	volume       = {10416},
	pages        = {689--706},
	publisher    = {Springer},
	year         = {2017},
	url          = {https://doi.org/10.1007/978-3-319-66158-2\_44},
	doi          = {10.1007/978-3-319-66158-2\_44},
	bibsource    = {dblp computer science bibliography, https://dblp.org}
}

@article{raghothaman2020provenance,
	author       = {Mukund Raghothaman and
	Jonathan Mendelson and
	David Zhao and
	Mayur Naik and
	Bernhard Scholz},
	title        = {Provenance-guided synthesis of {Datalog} programs},
	journal      = {Proc. {ACM} Program. Lang.},
	volume       = {4},
	number       = {{POPL}},
	pages        = {62:1--62:27},
	year         = {2020},
	url          = {https://doi.org/10.1145/3371130},
	doi          = {10.1145/3371130},
	bibsource    = {dblp computer science bibliography, https://dblp.org}
}

@article{bembenek2023smt2asp,
	author       = {Aaron Bembenek and
	Michael Greenberg and
	Stephen Chong},
	title        = {From {SMT} to {ASP:} Solver-Based Approaches to Solving {Datalog} Synthesis-as-Rule-Selection
	Problems},
	journal      = {Proc. {ACM} Program. Lang.},
	volume       = {7},
	number       = {{POPL}},
	pages        = {185--217},
	year         = {2023},
	url          = {https://doi.org/10.1145/3571200},
	doi          = {10.1145/3571200},
	bibsource    = {dblp computer science bibliography, https://dblp.org}
}

@inproceedings{mendelson2021gensynth,
	author       = {Jonathan Mendelson and
	Aaditya Naik and
	Mukund Raghothaman and
	Mayur Naik},
	title        = {{GENSYNTH:} Synthesizing {Datalog} Programs without Language Bias},
	booktitle    = {Thirty-Fifth {AAAI} Conference on Artificial Intelligence, {AAAI}
	2021, Thirty-Third Conference on Innovative Applications of Artificial
	Intelligence, {IAAI} 2021, The Eleventh Symposium on Educational Advances
	in Artificial Intelligence, {EAAI} 2021, Virtual Event, February 2-9,
	2021},
	pages        = {6444--6453},
	publisher    = {{AAAI} Press},
	year         = {2021},
	url          = {https://doi.org/10.1609/aaai.v35i7.16799},
	doi          = {10.1609/AAAI.V35I7.16799},
	bibsource    = {dblp computer science bibliography, https://dblp.org}
}

@inproceedings{si2019synthesizing,
	author       = {Xujie Si and
	Mukund Raghothaman and
	Kihong Heo and
	Mayur Naik},
	editor       = {Sarit Kraus},
	title        = {Synthesizing {Datalog} Programs using Numerical Relaxation},
	booktitle    = {Proceedings of the Twenty-Eighth International Joint Conference on
	Artificial Intelligence, {IJCAI} 2019, Macao, China, August 10-16,
	2019},
	pages        = {6117--6124},
	publisher    = {ijcai.org},
	year         = {2019},
	url          = {https://doi.org/10.24963/ijcai.2019/847},
	doi          = {10.24963/IJCAI.2019/847},
	bibsource    = {dblp computer science bibliography, https://dblp.org}
}

@inproceedings{si2018syntax,
	author       = {Xujie Si and
	Woosuk Lee and
	Richard Zhang and
	Aws Albarghouthi and
	Paraschos Koutris and
	Mayur Naik},
	editor       = {Gary T. Leavens and
	Alessandro Garcia and
	Corina S. Pasareanu},
	title        = {Syntax-guided synthesis of {Datalog} programs},
	booktitle    = {Proceedings of the 2018 {ACM} Joint Meeting on European Software Engineering
	Conference and Symposium on the Foundations of Software Engineering,
	{ESEC/SIGSOFT} {FSE} 2018, Lake Buena Vista, FL, USA, November 04-09,
	2018},
	pages        = {515--527},
	publisher    = {{ACM}},
	year         = {2018},
	url          = {https://doi.org/10.1145/3236024.3236034},
	doi          = {10.1145/3236024.3236034},
	bibsource    = {dblp computer science bibliography, https://dblp.org}
}

@inproceedings{scholz2016fast,
	author       = {Bernhard Scholz and
	Herbert Jordan and
	Pavle Subotic and
	Till Westmann},
	title        = {On fast large-scale program analysis in {Datalog}},
	booktitle    = {Proceedings of the 25th International Conference on Compiler Construction,
	{CC} 2016, Barcelona, Spain, March 12-18, 2016},
	pages        = {196--206},
	publisher    = {{ACM}},
	year         = {2016},
	url          = {https://doi.org/10.1145/2892208.2892226},
	doi          = {10.1145/2892208.2892226},
	bibsource    = {dblp computer science bibliography, https://dblp.org}
}

@inproceedings{jordan2016souffle,
	author       = {Herbert Jordan and
	Bernhard Scholz and
	Pavle Subotic},
	editor       = {Swarat Chaudhuri and
	Azadeh Farzan},
	title        = {Souffl{\'{e}}: On Synthesis of Program Analyzers},
	booktitle    = {Computer Aided Verification - 28th International Conference, {CAV}
	2016, Toronto, ON, Canada, July 17-23, 2016, Proceedings, Part {II}},
	series       = {Lecture Notes in Computer Science},
	pages        = {422--430},
	publisher    = {Springer},
	year         = {2016},
	url          = {https://doi.org/10.1007/978-3-319-41540-6\_23},
	doi          = {10.1007/978-3-319-41540-6\_23},
	bibsource    = {dblp computer science bibliography, https://dblp.org}
}

@inproceedings{thakkar2021example,
	author       = {Aalok Thakkar and
	Aaditya Naik and
	Nathaniel Sands and
	Rajeev Alur and
	Mayur Naik and
	Mukund Raghothaman},
	editor       = {Stephen N. Freund and
	Eran Yahav},
	title        = {Example-guided synthesis of relational queries},
	booktitle    = {{PLDI} '21: 42nd {ACM} {SIGPLAN} International Conference on Programming
	Language Design and Implementation, Virtual Event, Canada, June 20-25,
	2021},
	pages        = {1110--1125},
	publisher    = {{ACM}},
	year         = {2021},
	url          = {https://doi.org/10.1145/3453483.3454098},
	doi          = {10.1145/3453483.3454098},
	bibsource    = {dblp computer science bibliography, https://dblp.org}
}

@article{xie2024codefuse,
	author       = {Xiaoheng Xie and
	Gang Fan and
	Xiaojun Lin and
	Ang Zhou and
	Shijie Li and
	Xunjin Zheng and
	Yinan Liang and
	Yu Zhang and
	Na Yu and
	Haokun Li and
	Xinyu Chen and
	Yingzhuang Chen and
	Yi Zhen and
	Dejun Dong and
	Xianjin Fu and
	Jinzhou Su and
	Fuxiong Pan and
	Pengshuai Luo and
	Youzheng Feng and
	Ruoxiang Hu and
	Jing Fan and
	Jinguo Zhou and
	Xiao Xiao and
	Peng Di},
	title        = {{CodeFuse-Query}: {A} Data-Centric Static Code Analysis System for Large-Scale
	Organizations},
	journal      = {CoRR},
	volume       = {abs/2401.01571},
	year         = {2024},
	url          = {https://doi.org/10.48550/arXiv.2401.01571},
	doi          = {10.48550/ARXIV.2401.01571},
	eprinttype   = {arXiv},
	eprint       = {2401.01571},
	bibsource    = {dblp computer science bibliography, https://dblp.org}
}

@inproceedings{li2024deveval,
	author       = {Jia Li and
	Ge Li and
	Yunfei Zhao and
	Yongmin Li and
	Huanyu Liu and
	Hao Zhu and
	Lecheng Wang and
	Kaibo Liu and
	Zheng Fang and
	Lanshen Wang and
	Jiazheng Ding and
	Xuanming Zhang and
	Yuqi Zhu and
	Yihong Dong and
	Zhi Jin and
	Binhua Li and
	Fei Huang and
	Yongbin Li and
	Bin Gu and
	Mengfei Yang},
	editor       = {Lun{-}Wei Ku and
	Andre Martins and
	Vivek Srikumar},
	title        = {{DevEval}: {A} Manually-Annotated Code Generation Benchmark Aligned
	with Real-World Code Repositories},
	booktitle    = {Findings of the Association for Computational Linguistics, {ACL} 2024,
	Bangkok, Thailand and virtual meeting, August 11-16, 2024},
	pages        = {3603--3614},
	publisher    = {Association for Computational Linguistics},
	year         = {2024},
	url          = {https://doi.org/10.18653/v1/2024.findings-acl.214},
	doi          = {10.18653/V1/2024.FINDINGS-ACL.214},
	bibsource    = {dblp computer science bibliography, https://dblp.org}
}

@inproceedings{yu2024codereval,
	author       = {Hao Yu and
	Bo Shen and
	Dezhi Ran and
	Jiaxin Zhang and
	Qi Zhang and
	Yuchi Ma and
	Guangtai Liang and
	Ying Li and
	Qianxiang Wang and
	Tao Xie},
	title        = {{CoderEval}: {A} Benchmark of Pragmatic Code Generation with Generative
	Pre-trained Models},
	booktitle    = {Proceedings of the 46th {IEEE/ACM} International Conference on Software
	Engineering, {ICSE} 2024, Lisbon, Portugal, April 14-20, 2024},
	pages        = {37:1--37:12},
	publisher    = {{ACM}},
	year         = {2024},
	url          = {https://doi.org/10.1145/3597503.3623316},
	doi          = {10.1145/3597503.3623316},
	bibsource    = {dblp computer science bibliography, https://dblp.org}
}

@inproceedings{du2024evaluating,
	author       = {Xueying Du and
	Mingwei Liu and
	Kaixin Wang and
	Hanlin Wang and
	Junwei Liu and
	Yixuan Chen and
	Jiayi Feng and
	Chaofeng Sha and
	Xin Peng and
	Yiling Lou},
	title        = {Evaluating Large Language Models in Class-Level Code Generation},
	booktitle    = {Proceedings of the 46th {IEEE/ACM} International Conference on Software
	Engineering, {ICSE} 2024, Lisbon, Portugal, April 14-20, 2024},
	pages        = {81:1--81:13},
	publisher    = {{ACM}},
	year         = {2024},
	url          = {https://doi.org/10.1145/3597503.3639219},
	doi          = {10.1145/3597503.3639219},
	bibsource    = {dblp computer science bibliography, https://dblp.org}
}

@inproceedings{liu2023isurcode,
	author       = {Jiawei Liu and
	Chunqiu Steven Xia and
	Yuyao Wang and
	Lingming Zhang},
	editor       = {Alice Oh and
	Tristan Naumann and
	Amir Globerson and
	Kate Saenko and
	Moritz Hardt and
	Sergey Levine},
	title        = {Is Your Code Generated by {ChatGPT} Really Correct? Rigorous Evaluation
	of Large Language Models for Code Generation},
	booktitle    = {Advances in Neural Information Processing Systems 36: Annual Conference
	on Neural Information Processing Systems 2023, NeurIPS 2023, New Orleans,
	LA, USA, December 10 - 16, 2023},
	year         = {2023},
	url          = {http://papers.nips.cc/paper\_files/paper/2023/hash/43e9d647ccd3e4b7b5baab53f0368686-Abstract-Conference.html},
	bibsource    = {dblp computer science bibliography, https://dblp.org}
}

@article{chen2021evaluating,
	author       = {Mark Chen and
	Jerry Tworek and
	Heewoo Jun and
	Qiming Yuan and
	Henrique Pond{\'{e}} de Oliveira Pinto and
	Jared Kaplan and
	Harri Edwards and
	Yuri Burda and
	Nicholas Joseph and
	Greg Brockman and
	Alex Ray and
	Raul Puri and
	Gretchen Krueger and
	Michael Petrov and
	Heidy Khlaaf and
	Girish Sastry and
	Pamela Mishkin and
	Brooke Chan and
	Scott Gray and
	Nick Ryder and
	Mikhail Pavlov and
	Alethea Power and
	Lukasz Kaiser and
	Mohammad Bavarian and
	Clemens Winter and
	Philippe Tillet and
	Felipe Petroski Such and
	Dave Cummings and
	Matthias Plappert and
	Fotios Chantzis and
	Elizabeth Barnes and
	Ariel Herbert{-}Voss and
	William Hebgen Guss and
	Alex Nichol and
	Alex Paino and
	Nikolas Tezak and
	Jie Tang and
	Igor Babuschkin and
	Suchir Balaji and
	Shantanu Jain and
	William Saunders and
	Christopher Hesse and
	Andrew N. Carr and
	Jan Leike and
	Joshua Achiam and
	Vedant Misra and
	Evan Morikawa and
	Alec Radford and
	Matthew Knight and
	Miles Brundage and
	Mira Murati and
	Katie Mayer and
	Peter Welinder and
	Bob McGrew and
	Dario Amodei and
	Sam McCandlish and
	Ilya Sutskever and
	Wojciech Zaremba},
	title        = {Evaluating Large Language Models Trained on Code},
	journal      = {CoRR},
	volume       = {abs/2107.03374},
	year         = {2021},
	url          = {https://arxiv.org/abs/2107.03374},
	eprinttype    = {arXiv},
	eprint       = {2107.03374},
	bibsource    = {dblp computer science bibliography, https://dblp.org}
}

@inproceedings{jimenez2024swebench,
	author       = {Carlos E. Jimenez and
	John Yang and
	Alexander Wettig and
	Shunyu Yao and
	Kexin Pei and
	Ofir Press and
	Karthik R. Narasimhan},
	title        = {{SWE}-bench: Can Language Models Resolve Real-World {GitHub} Issues?},
	booktitle    = {The Twelfth International Conference on Learning Representations,
	{ICLR} 2024, Vienna, Austria, May 7-11, 2024},
	publisher    = {OpenReview.net},
	year         = {2024},
	url          = {https://openreview.net/forum?id=VTF8yNQM66},
	bibsource    = {OpenReview}
}

@inproceedings{zhang2025swebenchgoeslive,
	title={{SWE-bench} Goes Live!},
	author={Linghao Zhang and Shilin He and Chaoyun Zhang and Yu Kang and Bowen Li and Chengxing Xie and Junhao Wang and Maoquan Wang and Yufan Huang and Shengyu Fu and Elsie Nallipogu and Qingwei Lin and Yingnong Dang and Saravan Rajmohan and Dongmei Zhang},
	booktitle={Advances in Neural Information Processing Systems 38 (NeurIPS 2025)},
	pages={163347--163372},
	year={2025},
	doi={10.52202/085713-4923}
}

@inproceedings{jain2025livecodebench,
	author       = {Naman Jain and
	King Han and
	Alex Gu and
	Wen{-}Ding Li and
	Fanjia Yan and
	Tianjun Zhang and
	Sida Wang and
	Armando Solar{-}Lezama and
	Koushik Sen and
	Ion Stoica},
	title        = {{LiveCodeBench}: Holistic and Contamination Free Evaluation of Large Language Models for Code},
	booktitle    = {The Thirteenth International Conference on Learning Representations,
	{ICLR} 2025, Singapore, April 24-28, 2025},
	publisher    = {OpenReview.net},
	year         = {2025},
	url          = {https://proceedings.iclr.cc/paper_files/paper/2025/hash/94074dd5a072d28ff75a76dabed43767-Abstract-Conference.html},
	bibsource    = {ICLR 2025 Proceedings}
}

@inproceedings{barke2024hysynth,
	author       = {Shraddha Barke and
	Emmanuel Anaya Gonzalez and
	Saketh Ram Kasibatla and
	Taylor Berg{-}Kirkpatrick and
	Nadia Polikarpova},
	title        = {{HYSYNTH}: Context-Free {LLM} Approximation for Guiding Program Synthesis},
	booktitle    = {Advances in Neural Information Processing Systems 37: Annual Conference
	on Neural Information Processing Systems 2024, NeurIPS 2024, Vancouver,
	BC, Canada},
	year         = {2024},
	url          = {https://openreview.net/forum?id=5jt0ZSA6Co},
	bibsource    = {dblp computer science bibliography, https://dblp.org}
}

@inproceedings{li2025feabench,
	author       = {Wei Li and
	Xin Zhang and
	Zhongxin Guo and
	Shaoguang Mao and
	Wen Luo and
	Guangyue Peng and
	Yangyu Huang and
	Houfeng Wang and
	Scarlett Li},
	title        = {{FEA}-Bench: A Benchmark for Evaluating Repository-Level Code Generation
	for Feature Implementation},
	booktitle    = {Proceedings of the 63rd Annual Meeting of the Association for Computational
	Linguistics (Volume 1: Long Papers), {ACL} 2025, Vienna, Austria, July
	27 - August 1, 2025},
	pages        = {17160--17176},
	publisher    = {Association for Computational Linguistics},
	year         = {2025},
	url          = {https://aclanthology.org/2025.acl-long.839/},
	doi          = {10.18653/v1/2025.acl-long.839},
	bibsource    = {ACL Anthology}
}

@inproceedings{yu2018spider,
	author       = {Tao Yu and
	Rui Zhang and
	Kai Yang and
	Michihiro Yasunaga and
	Dongxu Wang and
	Zifan Li and
	James Ma and
	Irene Li and
	Qingning Yao and
	Shanelle Roman and
	Zilin Zhang and
	Dragomir R. Radev},
	title        = {{Spider}: A Large-Scale Human-Labeled Dataset for Complex and Cross-Domain
	Semantic Parsing and Text-to-{SQL} Task},
	booktitle    = {Proceedings of the 2018 Conference on Empirical Methods in Natural
	Language Processing, {EMNLP} 2018, Brussels, Belgium, October 31 -
	November 4, 2018},
	pages        = {3911--3921},
	publisher    = {Association for Computational Linguistics},
	year         = {2018},
	url          = {https://aclanthology.org/D18-1425/},
	doi          = {10.18653/v1/D18-1425},
	bibsource    = {dblp computer science bibliography, https://dblp.org}
}

@article{zhong2017seq2sql,
	author       = {Victor Zhong and
	Caiming Xiong and
	Richard Socher},
	title        = {{Seq2SQL}: Generating Structured Queries from Natural Language using
	Reinforcement Learning},
	journal      = {CoRR},
	volume       = {abs/1709.00103},
	year         = {2017},
	url          = {http://arxiv.org/abs/1709.00103},
	eprinttype   = {arXiv},
	eprint       = {1709.00103},
	bibsource    = {dblp computer science bibliography, https://dblp.org}
}

@inproceedings{li2023can,
	author       = {Jinyang Li and
	Binyuan Hui and
	Ge Qu and
	Jiaxi Yang and
	Binhua Li and
	Bowen Li and
	Bailin Wang and
	Bowen Qin and
	Ruiying Geng and
	Nan Huo and
	Xuanhe Zhou and
	Chenhao Ma and
	Guoliang Li and
	Kevin Chen{-}Chuan Chang and
	Fei Huang and
	Reynold Cheng and
	Yongbin Li},
	editor       = {Alice Oh and
	Tristan Naumann and
	Amir Globerson and
	Kate Saenko and
	Moritz Hardt and
	Sergey Levine},
	title        = {Can {LLM} Already Serve as {A} Database Interface? {A} BIg Bench for
	Large-Scale Database Grounded Text-to-{SQLs}},
	booktitle    = {Advances in Neural Information Processing Systems 36: Annual Conference
	on Neural Information Processing Systems 2023, NeurIPS 2023, New Orleans,
	LA, USA, December 10 - 16, 2023},
	year         = {2023},
	url          = {http://papers.nips.cc/paper\_files/paper/2023/hash/83fc8fab1710363050bbd1d4b8cc0021-Abstract-Datasets\_and\_Benchmarks.html},
	bibsource    = {dblp computer science bibliography, https://dblp.org}
}

@inproceedings{lei2025spider2,
	author       = {Fangyu Lei and
	Jixuan Chen and
	Yuxiao Ye and
	Ruisheng Cao and
	Dongchan Shin and
	Hongjin Su and
	Zhaoqing Suo and
	Hongcheng Gao and
	Wenjing Hu and
	Pengcheng Yin and
	Victor Zhong and
	Caiming Xiong and
	Ruoxi Sun and
	Qian Liu and
	Sida Wang and
	Tao Yu},
	title        = {{Spider} 2.0: Evaluating Language Models on Real-World Enterprise Text-to-{SQL}
	Workflows},
	booktitle    = {The Thirteenth International Conference on Learning Representations,
	{ICLR} 2025, Singapore, April 24-28, 2025},
	publisher    = {OpenReview.net},
	year         = {2025},
	url          = {https://openreview.net/forum?id=XmProj9cPs},
	bibsource    = {OpenReview}
}

@article{cropper2022ilp30,
	author       = {Andrew Cropper and
	Sebastijan Duman{\v{c}}i{\'{c}}},
	title        = {Inductive Logic Programming At 30: {A} New Introduction},
	journal      = {Journal of Artificial Intelligence Research},
	volume       = {74},
	pages        = {765--850},
	year         = {2022},
	doi          = {10.1613/jair.1.13507}
}

@article{muggleton2015meta,
	author       = {Stephen H. Muggleton and
	Dianhuan Lin and
	Alireza Tamaddoni{-}Nezhad},
	title        = {Meta-interpretive learning of higher-order dyadic {Datalog}: predicate
	invention revisited},
	journal      = {Machine Learning},
	volume       = {100},
	number       = {1},
	pages        = {49--73},
	year         = {2015},
	doi          = {10.1007/s10994-014-5471-y}
}

@article{cropper2021popper,
	author       = {Andrew Cropper and
	Rolf Morel},
	title        = {Learning programs by learning from failures},
	journal      = {Machine Learning},
	volume       = {110},
	number       = {4},
	pages        = {801--856},
	year         = {2021},
	doi          = {10.1007/s10994-020-05934-z}
}

@inproceedings{cerna2024generalisation,
	author       = {David M. Cerna and
	Andrew Cropper},
	title        = {Generalisation through Negation and Predicate Invention},
	booktitle    = {Proceedings of the {AAAI} Conference on Artificial Intelligence},
	volume       = {38},
	number       = {9},
	pages        = {10467--10475},
	year         = {2024},
	doi          = {10.1609/aaai.v38i9.28915}
}

@article{xu2026logicloc,
	author       = {Xiufeng Xu and
	Xiufeng Wu and
	Zejun Zhang and
	Yi Li},
	title        = {Neurosymbolic Repo-level Code Localization},
	journal      = {arXiv preprint arXiv:2604.16021},
	year         = {2026}
}

@misc{openai2026gpt56,
	author       = {{OpenAI}},
	title        = {{GPT-5.6} Preview System Card},
	year         = {2026},
	url          = {https://deploymentsafety.openai.com/gpt-5-6-preview},
	note         = {Accessed September 2026}
}

@misc{anthropic2026opus5,
	author       = {{Anthropic}},
	title        = {Introducing {Claude Opus 5}},
	year         = {2026},
	url          = {https://www.anthropic.com/news/claude-opus-5},
	note         = {Accessed September 2026}
}

@misc{google2026gemini38flash,
	author       = {{Google DeepMind}},
	title        = {{Gemini 3.8 Flash} Model Card},
	year         = {2026},
	url          = {https://deepmind.google/models/model-cards/gemini-3-8-flash/},
	note         = {Accessed November 2026}
}

@article{deepseek2026v4,
	author       = {{DeepSeek-AI}},
	title        = {{DeepSeek-V4}: Towards Highly Efficient Million-Token Context Intelligence},
	journal      = {arXiv preprint arXiv:2606.19348},
	year         = {2026}
}

@misc{openai2025codexcli,
	author       = {{OpenAI}},
	title        = {{Codex CLI}: A Lightweight Coding Agent That Runs in Your Terminal},
	year         = {2025},
	url          = {https://github.com/openai/codex},
	note         = {Accessed September 2026}
}

@misc{anthropic2025claudecode,
	author       = {{Anthropic}},
	title        = {{Claude Code}: An Agentic Coding Tool That Lives in Your Terminal},
	year         = {2025},
	url          = {https://github.com/anthropics/claude-code},
	note         = {Accessed September 2026}
}

@article{thakkar2023mobius,
  author = {Aalok Thakkar and Nathaniel Sands and George Petrou and Rajeev Alur and Mayur Naik and Mukund Raghothaman},
  title = {{Mobius}: Synthesizing Relational Queries with Recursive and Invented Predicates},
  journal = {Proceedings of the ACM on Programming Languages},
  volume = {7},
  number = {OOPSLA2},
  pages = {271:1--271:24},
  year = {2023},
  doi = {10.1145/3622847}
}

@inproceedings{zhong2020semantic,
  author = {Ruiqi Zhong and Tao Yu and Dan Klein},
  title = {Semantic Evaluation for Text-to-{SQL} with Distilled Test Suites},
  booktitle = {Proceedings of the 2020 Conference on Empirical Methods in Natural Language Processing (EMNLP)},
  pages = {396--411},
  year = {2020},
  publisher = {Association for Computational Linguistics},
  address = {Online},
  doi = {10.18653/v1/2020.emnlp-main.29},
  url = {https://aclanthology.org/2020.emnlp-main.29/}
}

\appendix

\section{Datalog Formalism and the Inductive Synthesis Problem}
\label{sec:pre}

\noindent Section~\ref{sec:prelim} states the fragment \bench uses. This appendix
fixes the notation, records the language features that fragment leaves out, and
states the inductive synthesis problem the task formulation instantiates.

\smallskip
\noindent \textbf{Syntax}.
A \emph{Datalog program} $P$ is a finite set of inference rules, each of the form
\[
A \horn B_1, B_2, \ldots, B_n
\]
where $A$ is the head and $\{B_1, \dots, B_n\}$ is the body. Head and body literals apply a predicate $p$ of fixed arity $\mathit{ar}(p)$ to a tuple of terms $(t_1, \dots, t_m)$, each term a constant $c$ or a variable $X$, and an EDB predicate never appears in a head. Beyond positive atoms, a body literal $B_i$ may be a negated atom $\neg B$, a comparison or arithmetic expression over terms, or an aggregate such as $\mathit{min}$ or $\mathit{sum}$ over an atom; the pure Horn fragment is the case in which every $B_i$ is a positive atom. Every variable in the head, in a negated atom, or in a comparison must also occur in a positive body atom (\emph{safety}).

\smallskip
\noindent \textbf{Semantics}.
The least fixed point of \S\ref{sec:prelim} is computed in practice by semi-naive evaluation, which avoids redundant derivations. For negation-free programs, $P$ is \emph{monotone} as a function from EDBs to IDBs: if $I \subseteq J$, then $P(I) \subseteq P(J)$. Programs with negation are evaluated stratum by stratum, and are well defined only under the stratification condition of \S\ref{sec:prelim}.

\begin{example}
	\label{exm:datalog}
	The running example of \S\ref{sec:prelim}, repeated here so that the synthesis problem below is self-contained. Over an input relation $\textit{Edge}$, the output relation $\textit{Reachable}$ holds for every pair connected by a path:
	\begin{align*}
		\textit{Reachable}(X, Y) &\horn \textit{Edge}(X, Y). \\
		\textit{Reachable}(X, Z) &\horn \textit{Edge}(X, Y), \textit{Reachable}(Y, Z).
	\end{align*}
	The second rule is recursive: an edge from $X$ to an intermediate node $Y$, followed by a path from $Y$ to $Z$, makes $Z$ reachable from $X$.
\end{example}

\begin{definition}
	[Datalog program synthesis]
	A Datalog program synthesis problem $S$ is a tuple $(R, I, O)$, where $R = (R_{in}, R_{out})$ is a pair of disjoint signatures of input and output relations, $I$ is a finite set of facts for the relations in $R_{in}$, and $O$ is a finite set of facts for the relations in $R_{out}$, partitioned into positive examples $O^{+}$ and negative examples $O^{-}$ with $O^{+} \cap O^{-} = \emptyset$.

	A solution is a Datalog program $P$ with $R_{in}(P) = R_{in}$ and $R_{out}(P) = R_{out}$ whose least fixed point $P(I)$ satisfies two conditions:

	\begin{itemize}
		\item $O^{+} \subseteq P(I)$: all positive examples are derived;
		\item $O^{-} \cap P(I) = \emptyset$: no negative example is derived.
	\end{itemize}
\end{definition}

\noindent This is the partial specification of inductive synthesis, which symbolic
synthesizers consume and on which negative exclusion (Appendix~\ref{app:metrics})
is built: any superset of $O^{+}$ that avoids $O^{-}$ is a solution. \bench grades
more strictly. Its primary metric requires the derived output to \emph{equal} the
expected output on every evaluation variant, a closed-world criterion under which
every tuple the reference does not derive counts as an error.

\begin{example}
	For the reachability problem of Example~\ref{exm:datalog}, a synthesis problem $S = (R, I, O)$ could be specified as follows:

	\begin{itemize}
		\item $R_{in} = \{\textit{Edge}/2\}$ and $R_{out} = \{\textit{Reachable}/2\}$, where the notation indicates arity;
		\item $I = \{\textit{Edge}(a, b), \textit{Edge}(b, c), \textit{Edge}(c, d)\}$;
		\item $O^{+} = \{\textit{Reachable}(a, b), \textit{Reachable}(b, c), \textit{Reachable}(a, c), \textit{Reachable}(a, d)\}$;
		\item $O^{-} = \{\textit{Reachable}(b, a), \textit{Reachable}(c, a), \textit{Reachable}(d, a),$ $\textit{Reachable}(c, b), \textit{Reachable}(d, b), \textit{Reachable}(d, c)\}$.
	\end{itemize}

	The positive examples specify paths that must exist, and the negative examples specify paths that must not (for instance, no path leads backward from $d$ to $a$). $O^{+}$ is deliberately a partial sample: the full closure also contains $\textit{Reachable}(b, d)$ and $\textit{Reachable}(c, d)$, which a partial specification leaves unconstrained. The recursive program of Example~\ref{exm:datalog} is a valid solution: it derives every positive example and no negative one.
\end{example}

The end-to-end pipeline is shown in Figure~\ref{fig:task}.

\section{Prompt Templates}
\label{app:prompts}

This appendix reports the prompts used for direct prompting and the feedback
message used by the agent harness. Angle brackets mark slots filled per task.

\noindent\textbf{System message.}
\begin{lstlisting}[breaklines=true]
You are an expert in Datalog program synthesis.
\end{lstlisting}

\noindent\textbf{User message.} The \emph{Domain Knowledge} block appears only
for tasks that carry a knowledge note, and the \emph{Few-shot IO Examples} block
holds the demonstration instance in the one-shot condition and reads
\texttt{(None)} otherwise. In the description condition, each relation line is
followed by its natural-language description.
\begin{lstlisting}[breaklines=true]
System Instruction:
You are an expert Datalog programmer specializing in the Souffle dialect. Write strict, safe, and efficient Datalog rules to satisfy the task.

Task Case: <case id>
Natural Language Query:
<question>

Input Relations:
- Signature: <relation signature>
  Description: <relation description>      (description condition only)
Output Relations:
- Signature: <relation signature>

Domain Knowledge:                           (tasks with a knowledge note only)
<knowledge note>

Synthesis Constraints:
1. Ensure all head variables appear in body rules (safety).
2. Use recursive rules when required by transitive semantics.
3. Output only final Datalog rules in one code block.
4. Do not include any comments in the generated Datalog code.

Few-shot IO Examples:
Example Variant demo/<instance>:
Input Tuples:
- <relation>:
  - (<value>, ...)
Output Tuples:
- <relation>:
  - (<value>, ...)

Final Output Format:
Return only one markdown code block in Souffle Datalog format.
```datalog
<rules>
```
\end{lstlisting}
The phrase ``body rules'' is reproduced verbatim from the evaluated prompt;
``body atoms'' or ``body literals'' would be the more precise terminology.

\noindent\textbf{Execution feedback.}
After each agent turn, the harness builds the following message from the
demonstration instance only; the evaluation variants are not executed until the
candidate is frozen.
\begin{lstlisting}[breaklines=true]
Your previous program was executed on the development I/O instance.
Previous Datalog program:
```datalog
<previous program>
```
Feedback:
<compile error, or bounded input and output-difference tuples>
Return a corrected Datalog program, only one ```datalog``` code block, no comments.
\end{lstlisting}

\section{Additional Experimental Details}
\label{app:model-details}

This appendix records the experimental details behind the results in
\S\ref{sec:result}: the evaluated systems and their request contracts, the
agent environment, the checks applied during annotation, and an analysis of
contamination from public upstream sources.

\subsection{Model Inventory}

\begin{table}[b]
	\centering
	\caption{Models, agents, and symbolic baselines evaluated on \bench, with the role of
	each in the design and the API identifier it was served under.}
	\label{tab:model-inventory}
	\resizebox{\textwidth}{!}{
		\begin{tabular}{@{}lllll@{}}
			\toprule
			\textbf{Setting} & \textbf{Provider} & \textbf{System} & \textbf{API identifier} & \textbf{Role in the design} \\
			\midrule
			Model & OpenAI & GPT~5.6~Sol & \texttt{gpt-5.6-sol} & Frontier breadth \\
			Model & Anthropic & Claude Opus~5 & \texttt{claude-opus-5} & Frontier breadth \\
			Model & Google & Gemini~3.8~Flash & \texttt{gemini-3.8-flash} & Frontier breadth \\
			Model & DeepSeek & DeepSeek~V4~Pro & \texttt{deepseek-v4-pro} & Frontier breadth; scale contrast vs.\ Flash \\
			Model & DeepSeek & DeepSeek~V4~FT (Flash, thinking) & \texttt{deepseek-v4-flash} & Reasoning ablation, thinking arm \\
			Model & DeepSeek & DeepSeek~V4~FNT (Flash, non-thinking) & \texttt{deepseek-v4-flash} & Reasoning ablation, non-thinking arm \\
			\midrule
			Agent & OpenAI & Codex CLI (Codex) / GPT~5.6~Sol & \texttt{codex} & Scaffold increment over the same model \\
			Agent & Anthropic & Claude Code (CC) / Claude Opus~5 & \texttt{claude} & Scaffold increment over the same model \\
			\midrule
			Symbolic & -- & GenSynth & -- & Genetic search; synthesis from examples \\
			Symbolic & -- & EGS & -- & Example-guided synthesis \\
			Symbolic & -- & ProSynth & -- & Provenance-guided CEGIS (\texttt{-{}-rule\_width}~$=2$) \\
			\bottomrule
		\end{tabular}
	}
\end{table}

Table~\ref{tab:model-inventory} lists the evaluated systems. Three properties of
this set are not apparent from the names alone. DeepSeek accounts for three of the
six models because it exposes a reasoning toggle on an otherwise identical model:
both Flash arms request the same API model, \texttt{deepseek-v4-flash}, and
differ only in an explicit \texttt{thinking} field, which makes the pair a
controlled ablation rather than a comparison across families (provenance in
Appendix~\ref{app:provenance}). DeepSeek~V4~Pro and V4~FT both
run at an explicitly low reasoning effort, so the Pro--Flash comparison holds
reasoning mode and effort fixed while varying scale. Finally, Google appears only
through its Flash tier because its Pro line has not been updated past an earlier
generation; including a generation-old Pro model would have mixed generations in
the frontier set.

\subsection{Request Contracts and Runtime}
\label{app:contracts}

All models receive the same system message and user-message template
(Appendix~\ref{app:prompts}) with no model-specific prompt tuning, and one completion per task in each run. Request contracts are
recorded per case rather than left to SDK defaults. All six models are requested with a $32{,}768$-token output
budget. The DeepSeek entries request temperature~$0$. GPT~5.6~Sol runs at
temperature~$1$, while Claude Opus~5 and Gemini are called without sampling overrides.
GPT, Claude, and Gemini use a $300$-second request timeout, and the three DeepSeek
entries stream responses with a $600$-second outer bound and at most two
transport attempts. V4~Pro and V4~FT additionally request low
reasoning effort; the other models omit reasoning-effort overrides. A response that exhausts the budget without a program counts as a failure. Generated
programs are compiled and executed with Souffl{\'e}~2.5 under a wall-clock timeout
of $30$ seconds, and a task counts as solved only if its program compiles and its
output matches the oracle on every evaluation variant.

\noindent\textbf{Agent environment.} The agents are invoked through the
command-line entry points in Table~\ref{tab:model-inventory}, inside a container
that supplies the same boundary to both. Only the per-task scratch directory is
mounted, so the reference programs are absent rather than merely unreadable. The
container's network is internal, and its only route out is a proxy whose allowlist
holds the model endpoint alone, so each CLI reaches its model and nothing else. We
verify four properties inside the container rather than trusting the
configuration: the reference program is unreadable by absolute path, a public URL
fails to resolve, the model endpoint responds, and Souffl{\'e} compiles and runs.
Each check is a single reproducible command, released with the image definitions.

Because agent tooling is versioned far more loosely than model weights, the image
pins the CLI builds and the Souffl{\'e} version, and each run records the image
digest together with the versions read out of the image itself. Both agent
settings ran in image \texttt{sha256:7f0230b6e00d}, which carries Souffl{\'e}~2.5 at
a $64$-bit word size, \texttt{codex-cli 0.153.4}, and CC
\texttt{2.1.261}. All $816$ recorded cases agree on these values, so every cell of
the agent grid was graded by the same compiler and run by the same client builds.
The image digest, rather than any version string, identifies what ran.

\subsection{Settings and Reproducibility}
\label{app:reproducibility}
The benchmark, prompts, evaluation harness, scoring scripts, and the complete
experiment run matrix are available in the code repository (\repourl). Experiments were run on a server with two AMD
EPYC~9754 processors ($256$ physical cores, $512$ hardware threads) and approximately $1.5$~TiB of RAM, running Ubuntu~22.04. Individual tools retain the concurrency
limits declared by the harness; in particular, GenSynth uses eight worker
processes. All model outputs come from hosted endpoints, including DeepSeek~V4~Flash,
whose open-weight revision we did not deploy locally, so they reproduce only as
far as the providers continue to serve the recorded builds
(Appendix~\ref{app:provenance}).

\noindent\textbf{Repeated runs.} Each configuration has three reported runs. The prompts and request settings are
fixed across reported runs, as is the agent container. We summarize run-level
scores by their mean and sample standard deviation. For inference, each task's
outcome is its mean over its reported runs; bootstrap resampling and paired
comparisons operate on tasks, not individual programs. Run-level standard
deviations describe variability across runs; task-level intervals describe
uncertainty over tasks. Program counts pool the reported runs.

\subsection{Setup Details Omitted from the Main Text}
\label{app:setup-details}

\noindent\textbf{Demonstration instances.} The main one-shot condition uses the
single demonstration instance supplied with each task.

\noindent\textbf{A second example.} We wrote a second instance for $40$ tasks
sampled with a fixed seed across domain and complexity tier. Each satisfies the
encoded input constraints, duplicates no demonstration, evaluation, or
additional-test input, and takes its expected output from the reference.
The six models receive relation signatures and two examples, with three reported
runs. We compare one and two examples on the same tasks and
runs. Paired differences are averaged over runs within each task and equally
over models, and differences are computed before rounding.
Mean EX changes by $+1.7$ points ($95\%$ CI $[-1.5,4.9]$); no model's change
has an interval excluding zero (Table~\ref{tab:second-example}).

\begin{table}[htbp]
	\centering
	\small
	\caption{EX (\%) with one or two examples under relation signatures on the
	$40$ sampled tasks, averaged over \nruns{} runs. Differences are paired by task;
	$95\%$ intervals and sign-flip $p$-values are shown.}
	\label{tab:second-example}
	\begin{tabular}{lccccc}
		\toprule
		Model & One example & Two examples & Difference & $95\%$ CI & $p$ \\
		\midrule
		GPT~5.6~Sol & 67.5 & 66.7 & $-0.8$ & $[-9.2,8.3]$ & $0.884$ \\
		Claude Opus~5 & 76.7 & 78.3 & $+1.7$ & $[-5.0,9.2]$ & $0.788$ \\
		Gemini~3.8~Flash & 75.8 & 78.3 & $+2.5$ & $[-3.3,9.2]$ & $0.650$ \\
		DeepSeek~V4~Pro & 55.0 & 60.8 & $+5.8$ & $[-1.7,14.2]$ & $0.251$ \\
		DeepSeek~V4~FT & 56.7 & 57.5 & $+0.8$ & $[-4.2,6.7]$ & $0.947$ \\
		DeepSeek~V4~FNT & 49.2 & 49.2 & $0.0$ & $[-7.5,6.7]$ & $1.000$ \\
		\midrule
		Mean of six models & 63.5 & 65.1 & $+1.7$ & $[-1.5,4.9]$ & $0.347$ \\
		\bottomrule
	\end{tabular}
\end{table}

\noindent\textbf{Knowledge notes supply vocabulary, not structure.} A model that does not know
what escaping means has not failed at recursion or at predicate invention; it has
failed at vocabulary, which is not the capability under study. The note therefore
states what terms and values mean and never how to structure the program, since a
note that named the auxiliary predicates would hand over the part of the task
the benchmark measures.

\noindent\textbf{Enforcement of the agent restrictions.} The turn budget is a
harness argument, and the tool restriction is a per-CLI policy: for Codex, web
search is opt-in through a flag the harness never passes; for CC, an
allowlist names only the local file and shell tools. An agent with no policy
defined refuses to run and reports that it is unconstrained, rather than
proceeding silently. Retrieval is disabled because every upstream artifact is
public together with its reference program (Appendix~\ref{app:contamination}),
so an agent able to search would be able to look the answer up; an agent whose
retrieval cannot be disabled from its client is therefore outside the scope of
this setting.

\noindent\textbf{ProSynth's hypothesis space.} Raising \texttt{-{}-rule\_width}
enlarges the candidate space steeply---roughly $78$ candidates at width $2$
against about $2{,}000$ at width $3$ for a simple task---and width $3$ typically
exceeds the time budget during preparation. We therefore report ProSynth at its
default width and treat the bound as a scaling property of rule-selection
approaches rather than a configuration artifact.

\subsection{Annotation Quality Control}
\label{app:annotation-quality}

During construction, the authors compared questions and relation descriptions
with reference programs, checked argument roles, and re-executed the references
against the retained instances. Three co-authors with Datalog or logic-programming
experience discussed annotation choices. This curation was separate from the
independent review below.

\begin{samepage}
\noindent\textbf{Independent review and follow-up.} Three non-author reviewers
who had not seen the reference programs or expected outputs independently
assessed $24$ unmodified specifications: two per cell of four domains and three
complexity tiers. Six hidden controls omitted a key sentence. Reviewers received
the question, signatures, relation descriptions, and any existing knowledge note;
references, expected outputs, input--output examples, and control identities were
withheld. They recorded their interpretation and required assumptions.\par
\end{samepage}

All three judged $11$ of the $24$ specifications self-contained. Two further
specifications received at least one domain-knowledge-dependence judgment and no
underdetermination judgment; the remaining $11$ were judged underdetermined by
at least one reviewer. The three reviewers labelled $6/6$, $6/6$, and $3/6$
deletion controls underdetermined. These judgments are quality checks, not
proofs of sufficiency.

The same reviewers completed a targeted follow-up on $13$ items using paired
interpretations, small inputs, and conditional output predictions. Follow-up
responses were retained separately from the initial responses and did not
replace the original labels.

\subsection{Contamination Analysis}
\label{app:contamination}

Every task in \bench derives from a public source---Souffl{\'e},
EGS, GenSynth, or CodeFuse-Query---so no part of the benchmark can be claimed to
be unexposed.
The relevant question is therefore not whether the upstream material
is public, but what having seen it is worth. All figures below are produced by
the released contamination scan.

Three properties of the construction limit that value. Relations were renamed and
their argument types standardized during normalization
(\S\ref{sec:construction-phases}); input facts were regenerated rather than
reused; and the schema a solver must target is supplied in the prompt. A model
reproducing an upstream program verbatim would therefore emit relation names the
task does not declare and fail before its logic was evaluated. Passing requires
mapping onto the schema actually given, which surface recall does not supply.

Renaming does not block recall of the algorithm, and here the two ends of the task
distribution differ. For standard constructions such as transitive closure,
reachability, and same-generation, knowing the shape of the program is ordinary
competence rather than contamination, and a benchmark that penalized it would
measure the wrong thing. For idiosyncratic targets, such as a particular
context-sensitivity choice in a pointer analysis or a specific parsing chart,
recall of the algorithm is closer to genuine exposure, and we cannot exclude it.

We do not partition the benchmark into exposed and clean subsets. Stratifying
tasks by whether their upstream source published a Datalog solution would rest on
matching names or signatures, and because normalization renamed relations to
different degrees across tasks, such a match would measure how thoroughly a task
was rewritten rather than whether a model has seen it.

One distinction does survive and is the closest the benchmark comes to a control.
The tasks drawn from CodeFuse-Query were published as G\"odel programs, not
Datalog; their Datalog reference programs and fact sets were written by us, and we
have not traced them to any of the other three sources. For those $27$ tasks the
target artifact in the language under test never existed publicly, whereas the
remaining $109$ trace to sources that did publish Datalog. The underlying logic was
public in both cases, so this is a difference in kind rather than a clean
partition, but a model carried by recall of Datalog text should do relatively
worse where no Datalog text existed. Averaged over the $24$ direct-prompting
cells and the \nruns{} runs, exact match on the CodeFuse-Query tasks is $59.1\%$ against $56.5\%$ on the
rest, a gap of $2.5$ points ($95\%$ CI $[-10.9,15.4]$) in the direction opposite to what contamination would
predict, and the CodeFuse-Query subset is ahead in $17$ of the $24$ cells. With $27$ tasks and a difference in
domain as well as provenance between the groups, this is weak evidence either way.

A more direct check looks for recall that surfaces as a name. A model reciting an
upstream program could emit a predicate name that occurs in the upstream artifact
but is absent from the provided schema. Over the $10{,}608$ programs of the reported configurations---the $24$
Direct cells and the two agents' signature runs, over the \nruns{} runs---the scan flags $1{,}831$
such occurrences. Many flags are ordinary names for auxiliary concepts;
the scan can also match non-code text. Common examples include
\texttt{Reach} and \texttt{Reachable} in strongly-connected-component and
disconnectedness tasks, \texttt{Ancestor} in class-hierarchy tasks, \texttt{State}
in automaton tasks. Two counts support this judgment. First, $638$ of the $1{,}831$
are names that our own annotators chose independently as auxiliary predicates in
the reference program for the same task; when the upstream author and our
annotator reach for the same word, a model reaching for it too is converging on
the obvious name. Second, $1{,}227$ of the $1{,}831$ appear as auxiliary predicates
somewhere in our reference programs, so they belong to the benchmark's own
working vocabulary. The $604$ outside that reference vocabulary include generic domain terms,
led by \texttt{Reach} ($307$), \texttt{Ancestor} ($67$), and \texttt{PointsTo}
($46$), which name the concept the task asks for rather than any upstream
program's idiosyncratic choice. The scan is in the repository and takes the
upstream sources as a path argument; all four are public
(\S\ref{sec:construction-phases}) and we redistribute none of them. Its limitation is that it detects only recall that surfaces as a
name: a model reproducing an upstream program's \emph{structure} under the
schema's own names would pass it unnoticed.

The scan cannot exclude structural recall under renamed predicates, so we treat
it as a contamination diagnostic rather than a guarantee. The benchmark's schema
normalization and held-out execution still require a recalled algorithm to be
adapted to the task interface and behaviorally correct.

\section{Benchmark Construction in Detail}
\label{app:construction}

\bench was built in four phases by authors with computer-science backgrounds and prior Datalog programming experience. Section~\ref{sec:construction-phases} states what each phase guarantees; this appendix records how, and Figure~\ref{fig:bench_con} summarizes the four phases.

\noindent \textbf{Phase 1: Query and fact collection.}
To curate a reproducible benchmark, we collect Datalog queries, along with input--output facts for validation, from existing Datalog synthesis artifacts and domain-specific tools using query languages. We sample cases from four sources:

\begin{itemize}
	\item Souffl{\'e} (UPL-1.0), a Datalog solver that compiles Datalog into native parallel C++ programs.
	\item GenSynth (no license declared; referenced as an unmodified submodule), a tool synthesizing Datalog programs without language bias.
	\item EGS (MIT), a tool that exploits example patterns to accelerate Datalog synthesis.
	\item CodeFuse-Query (Apache-2.0), a tool that expresses complex code-analysis tasks in the Gödel logical programming language.
\end{itemize}

\begin{table}[htbp]
\centering
\small
\caption{Candidates and retained tasks by source. Souffl{\'e} includes six
documentation examples among its candidates, of which the reaching-definitions
example (DFA) was retained.}
\label{tab:source-retention}
\begin{tabular}{lrrrr}
\toprule
Source & Candidates & Sanitized & Before audit & Final \\
\midrule
Souffl{\'e} & 123 & 85 & 72 & 68 \\
EGS & 86 & 50 & 25 & 25 \\
GenSynth & 47 & 17 & 16 & 16 \\
CodeFuse-Query & 31 & 31 & 27 & 27 \\
\midrule
Total & 287 & 183 & 140 & 136 \\
\bottomrule
\end{tabular}
\end{table}

Table~\ref{tab:source-retention} records the candidates and retained tasks at each
stage. Sanitization removes examples that depend on complex types (records or
abstract data types) or non-core language components (components,
\texttt{\#include} directives, or user-defined functors). Other exclusions cover
cross-source duplicates, engine tests, and simple SQL-style selection or join tasks.
Variant families are consolidated rather than counted as distinct tasks.
Among the documentation examples, the targets in \texttt{path} and \texttt{scc}
are covered by Path and SCC, while \texttt{cfg} and \texttt{sgen} were consolidated
with existing reachability and same-generation variants.
\texttt{VarPointsTo} was excluded as a scope choice: it is a simplified
implementation within the pointer-analysis family already represented in the
benchmark, rather than a duplicate of a retained program.
The specification audit removes one duplicate (Clique, which repeats SCC) and
three tasks whose specifications could not be made determinate without
restating their implementations (PointerReach, Puzzle, and Sequences).

For cases that provide relation signatures and input--output pairs but no ground-truth program, we manually author reference implementations. We also translate queries written in Gödel into standard Datalog, extract relation signatures, and generate representative input--output examples.

Cross-source copies are retained once and near-identical variants consolidated.
Selected distinct algorithms for the same problem remain separate; multi-output
programs may yield several tasks.

The resulting benchmark does not cover large industrial analyses with hundreds
of rules and millions of facts, complex types, or modular language extensions. Its scope is functional synthesis of
individual queries rather than full analysis systems or execution-performance
optimization.

\smallskip
\noindent \textbf{Phase 2: Data normalization.}
Each case is normalized into a structured format with several fields:
\begin{itemize}
	\item \emph{Case ID}: a unique identifier for the Datalog program synthesis task.
	\item \emph{Category information}: a two-level taxonomy of task domains. For example, a case may be classified as \textit{graph analytics} and further specified as \textit{reachability or connectivity}.
	\item \emph{Natural-language question}: a specification of the input-to-output transformation.
	\item \emph{Relation signatures and descriptions}: standardized names and semantic descriptions for each input and output relation.
\end{itemize}

Each case is originally assigned a ground-truth Datalog program and a set of input--output facts. To strengthen validation, each case includes additional input--output variants, selected for the reference-program mutants they distinguish rather than generated by blind perturbation (\S\ref{sec:oracle-strength}); the benchmark ships $251$ evaluation variants across its $136$ tasks, plus one demonstration instance per task that is never graded.

\smallskip
\noindent\emph{Storage layout.}
The three components of a task are stored separately rather than bundled into one record. A single JSON file holds the natural language question, the relation signatures, and their descriptions; the reference program lives in its own \texttt{.dl} file carrying the typed declarations, input and output directives, and rules; and each input--output variant occupies its own directory of fact and expected-output files. Keeping the reference program as a standalone executable artifact rather than as a string embedded in JSON is what allows the structural analysis of \S\ref{sec:stats} and the quality checks of Appendix~\ref{app:audit} to run directly over it, and it lets the Soufflé harness execute a reference program without any extraction step that could itself introduce discrepancies.

\smallskip
\noindent\emph{Type discipline.}
Soufflé requires every relation argument to be typed, and the choice interacts with semantics: \texttt{number} enables arithmetic and comparison but imposes numeric ordering, while \texttt{symbol} treats values as opaque identifiers. Because the identity of a node, variable, or heap object carries no numeric meaning, we prefer \texttt{symbol} wherever it is safe, and retain \texttt{number} only where a task genuinely computes over quantities---arithmetic, aggregation, or order-sensitive semantics such as interval overlap and weighted distance. Retyping was applied per file and accepted only if the reference program still matched its oracle exactly on all of its evaluation variants afterward, so no type change silently altered a task's meaning. In the released benchmark, $86$ reference programs use \texttt{symbol} exclusively, $17$ use \texttt{number} exclusively, and $33$ mix the two.

\smallskip
\noindent \textbf{Phase 3: Oracle verification.}

We validate the benchmark through automated testing with Soufflé~\citep{jordan2016souffle}. A task passes verification only if its reference program: (1) completes execution within a predefined time bound ($\le 30s$), and (2) produces outputs that match the expected results across all input--output examples. All $136$ tasks satisfy both conditions in the released benchmark.

This establishes that a reference program agrees with its own oracle. That is
well-formedness, and it is weaker than it looks: a program can agree with its
oracle while carrying a rule that never influences an output, or while exporting
a relation that no expected facts grade. Two standing checks close those gaps and
run in continuous integration rather than once at construction time.
\emph{Rule reachability} reports rules whose head relation no output depends on;
\emph{orphan-output detection} reports relations the reference exports that carry
no expected facts, and so are graded on nothing. Both report zero on the released
benchmark: no reference program carries a rule that no output depends on, and no
reference exports a relation that nothing grades. The structural statistics of
\S\ref{sec:stats} are computed on this released state.

Well-formedness is not the same as discrimination, however. A benchmark can be
perfectly self-consistent and still fail to separate a correct program from a
wrong one---in the limit, a task whose expected outputs were all empty would pass
every check above. Section~\ref{sec:validation} measures that second property.

\smallskip
\noindent \textbf{Phase 4: Question and relation annotation.}
Each case is augmented with a natural language specification and standardized relation annotations. Annotators write each question by hand to be natural and unambiguous, under three requirements:
\begin{itemize}
	\item \emph{Interface grounding}: the question states the intended input-to-output transformation in terms of the declared input and output relations, so that every relation the task is graded on is named in the specification.
	\item \emph{Quantification precision}: annotators use precise quantifiers (``at least one'', ``exactly'', ``all'') to reduce ambiguity.
	\item \emph{Domain terminology}: specifications use domain-specific terms that are consistent with the assigned taxonomy.
\end{itemize}

Relation names and parameters are standardized with manually curated semantic descriptions. All questions and relation descriptions are written by the authors; no part of the released specification text is machine-generated. During construction, annotators discussed interpretation choices and revised the question and relation descriptions. Appendix~\ref{app:annotation-quality} describes the author checks and the subsequent independent review.

\section{Benchmark Statistics in Detail}
\label{app:stats}

Section~\ref{sec:stats} reports the figures the results are read against; this
appendix gives the structural profile, data scale, domain coverage, the two
difficulty axes, and the structural profile by domain in full.

\begin{figure}[t]
	\centering
	\includegraphics[width=0.75\linewidth]{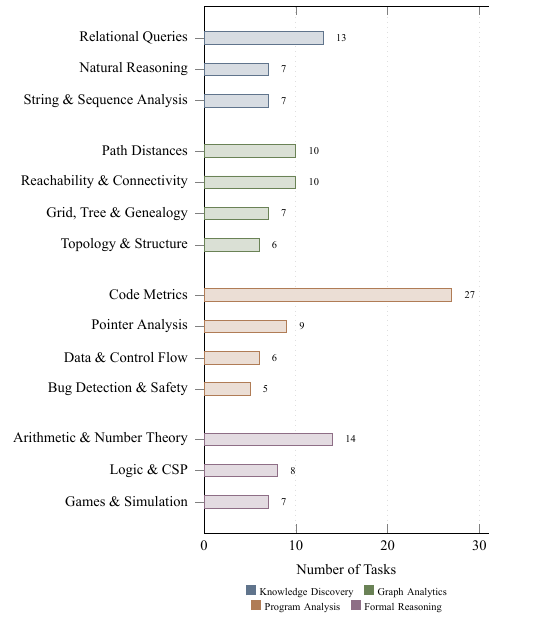}
	\caption{Tasks per sub-category, grouped by reasoning domain.}
	\label{fig:distribu}
\end{figure}

\noindent\textbf{Structural complexity.}
Table~\ref{tab:complex} reports the structural profile of the reference programs, extracted automatically from each reference program's predicate dependency graph. Reference programs average $5.1$ rules, with a median of $3$ and a maximum of $57$.

Recursion is prevalent, as it is the capability text-to-SQL benchmarks rarely exercise: $78$ of the $136$ tasks ($57.4\%$) are recursive, where a task counts as recursive if its predicate dependency graph contains a cycle, covering self-recursion and mutual recursion alike. Recursion is not merely present but structured: the largest strongly connected component in a single task spans $9$ mutually recursive predicates, and one task contains $4$ distinct recursive clusters.

Beyond recursion, \bench exercises the two constructs that impose Datalog's stratification obligations. $34$ tasks use negation and $22$ use aggregation, both of which require the solver to evaluate the program in dependency order rather than as a flat set of implications. All $136$ reference programs are stratified, since Souffl{\'e} compiles every one of them, and require $1.5$ strata on average, up to $5$.

\smallskip
\noindent\textbf{Data scale.}
Tasks involve an average of $2.9$ input and $1.5$ output relations. On each task's first evaluation variant, the fact-count distribution is heavy-tailed: the mean is $82$ input and $1{,}517$ output tuples, but the medians are only $10$ and $5$, respectively, and the trimmed means excluding the five largest cases are $12.5$ and $12.3$, as seen in Table~\ref{tab:complex}. We retain the long tail because some realistic Datalog workloads such as pointer analysis and large-graph reachability only exhibit scaling issues at non-trivial input sizes; small inputs alone would allow trivially incorrect programs to pass.

\smallskip
\noindent\textbf{Domain coverage.}
Figure~\ref{fig:distribu} shows how tasks distribute across four reasoning domains, each chosen to exercise a different aspect of declarative synthesis:
\begin{itemize}
	\item \textit{Program analysis} ($n=47$) covers code metrics and inter-procedural call-graph reachability ($n=27$), pointer analysis, including call-site-, object-, and type-sensitive variants ($n=9$), dataflow and control-flow analysis ($n=6$), and bug detection and safety properties ($n=5$). It is the historical core application of Datalog. Structurally, it is more heterogeneous than the label suggests: $43\%$ of its tasks are recursive, since many code-metric queries are aggregations over a call graph rather than fixed-point computations, and its input schemas are among the widest in the benchmark (up to $20$ input relations).
	\item \textit{Graph analytics} ($n=33$) targets recursive fixed-point algorithms: reachability and connectivity ($n=10$), path-distance and algebraic path problems ($n=10$), graph topology and structural analysis ($n=6$), and specialized structures such as grids, trees, and genealogies ($n=7$).
	\item \textit{Formal reasoning} ($n=29$) tests Datalog's expressive limits on arithmetic and number theory ($n=14$), constraint satisfaction and Boolean satisfiability, including 2-SAT and Circuit-SAT ($n=8$), and game simulation ($n=7$). This domain emphasizes encoding formal constraints in a relational language.
	\item \textit{Knowledge discovery} ($n=27$) covers relational queries with projection, selection, and aggregation ($n=13$), string and sequence analysis such as edit distance ($n=7$), and natural reasoning tasks ($n=7$). It serves as the structural baseline of the benchmark, with the lowest recursion rate ($37\%$) and, more distinctively, far less predicate invention than any other domain ($0.6$ auxiliary predicates per task against $1.6$ elsewhere): these tasks are largely expressible in terms of the relations the schema already provides. It also contains the widest input schema in the benchmark ($30$ relations).
\end{itemize}

\smallskip
\noindent\textbf{Two axes of difficulty.}
A benchmark on which everything is uniformly hard reports one number and explains
little, so we also score task difficulty, along two axes that we keep separate. \emph{Program complexity} asks how hard the target
program is to write once one knows what to write (rule count, largest recursive
component, strata, body width, largest arity); \emph{specification load} asks how much
the statement leaves for the solver to supply (auxiliary predicates, input
relations, question brevity). A single composite would confound them, and the
second is the quantity under dispute: if low scores merely track complex target
programs, that is an unremarkable fact about difficulty, whereas if they partly
track thin specifications then the benchmark is measuring specification quality
rather than model capability. Each axis is the mean percentile rank of its
metrics---ranks rather than a weighted sum, since a rule count, an arity, and a
word count share no unit---split into equal tiers (Table~\ref{tab:difficulty}).

The two axes correlate at $\rho = +0.333$, which is the informative outcome: near
$1$ and specification load would add nothing beyond complexity; near $0$ and
pairing them would be arbitrary; a moderate value says they overlap yet remain
separable enough to vary one independently. The domains sit differently on the
two (Table~\ref{tab:difficulty-domain}): formal reasoning writes the longest programs of
any domain---a median of five rules---from the lightest specifications, with over
half its tasks in the easiest specification-load tier, because a question that
states a mathematical condition states it completely and leaves only the encoding
to be found. Program analysis is the mirror image, carrying the heaviest
specification load---$26$ of its $47$ tasks in the hardest tier---on programs of
median length three, since what it asks for is named in a phrase, an escaping
object or a reaching definition, whose expansion into relations is the task. A
merged score would have recorded the two as comparably difficult. Both axes are
structural proxies rather than observed difficulty. Averaged over the $24$ Direct cells, tasks needing a domain fact the
schema cannot carry are solved $37.9\%$ of the time against $61.3\%$ for the
rest ($45.6\%$ for the $15$ that carry a knowledge note, $26.5\%$ for the $10$ that
do not), a gap of $23.4$ points that is strongly confounded by program complexity:
the L1-minus-L2 gap is $-1.3$ points in the easiest complexity bin ($5$ labeled
tasks), $+39.7$ in the middle bin ($6$), and $+6.5$ in the hardest ($14$). With
so few labeled tasks per bin, the label's own effect cannot be separated from
program complexity. We therefore report the label as a description of how the benchmark is
composed and decline to present either axis as a calibrated scale; separating
the two would need tasks that vary in one while holding the other fixed, which
this library does not contain in usable numbers.

\begin{table}[t]
	\centering
	\small
	\caption{Difficulty tiers along two axes, as the median of each raw metric within a
	tier. Max SCC has median $1$ in every tier because singleton SCCs dominate,
	including in non-recursive programs.}
	\label{tab:difficulty}
	\resizebox{0.98\textwidth}{!}{%
	\begin{tabular}{lccccc@{\hskip 2em}lcccc}
		\toprule
		\multicolumn{6}{c}{\textbf{Program complexity}} & \multicolumn{5}{c}{\textbf{Specification load}} \\
		\cmidrule(lr){1-6}\cmidrule(lr){7-11}
		Tier & $n$ & Rules & Max SCC & Strata & Largest arity & Tier & $n$ & Invented & Inputs & NL words \\
		\midrule
		Easy   & 46 & 2 & 1 & 1 & 2 & Easy   & 46 & 0 & 2 & 25 \\
		Medium & 45 & 3 & 1 & 1 & 3 & Medium & 45 & 0 & 2 & 14 \\
		Hard   & 45 & 6 & 1 & 2 & 4 & Hard   & 45 & 2 & 4 & 13 \\
		\bottomrule
	\end{tabular}
	}
\end{table}

\smallskip
\begin{table}[t]
	\centering
	\small
	\caption{Difficulty tier distribution by domain along each axis.}
	\label{tab:difficulty-domain}
	\begin{tabular}{lccc@{\hskip 2em}ccc}
		\toprule
		& \multicolumn{3}{c}{\textbf{Program complexity}} & \multicolumn{3}{c}{\textbf{Specification load}} \\
		\cmidrule(lr){2-4}\cmidrule(lr){5-7}
		Domain & Easy & Medium & Hard & Easy & Medium & Hard \\
		\midrule
		Program analysis    & 10 & 19 & 18 & \phantom{0}7 & 14 & 26 \\
		Graph analytics     & 15 & 10 & \phantom{0}8 & 13 & 15 & \phantom{0}5 \\
		Formal reasoning    & \phantom{0}9 & \phantom{0}7 & 13 & 15 & \phantom{0}8 & \phantom{0}6 \\
		Knowledge discovery & 12 & \phantom{0}9 & \phantom{0}6 & 11 & \phantom{0}8 & \phantom{0}8 \\
		\bottomrule
	\end{tabular}
\end{table}

\smallskip
\noindent\textbf{Structural differences between domains.}
Domain labels are useful only if they track something measurable, so we check them against the structural profile (Table~\ref{tab:domain-structure}). They separate along two axes, and not the one the labels suggest. \emph{Recursion} is highest in formal reasoning ($79.3\%$) and graph analytics ($75.8\%$), not in program analysis ($42.6\%$)---encoding a constraint problem as a fixed point turns out to demand recursion more reliably than analyzing a call graph does. \emph{Predicate invention} separates knowledge discovery ($0.6$ per task) from everything else ($1.6$ apiece). Graph analytics is the most structurally demanding on both counts that concern the solver: it has the largest recursive components (up to $9$ mutually recursive predicates), the deepest stratification ($1.8$ strata on average), and the heaviest aggregation use ($11$ of its $33$ tasks).

Two consequences matter for how the results should be read. First, the domains are not interchangeable samples of one capability, which is what makes a per-domain breakdown informative. Second, the labels are domain-of-origin labels, not difficulty labels: program analysis is the least recursive domain but carries the largest input schemas, so a model may fail there for reasons unrelated to fixed-point reasoning. We therefore report results sliced by structural property and by domain (\S\ref{sec:failure}).

\begin{table}[t]
	\centering
	\small
	\caption{Structural profile by domain.}
	\label{tab:domain-structure}
	\resizebox{0.98\textwidth}{!}{%
	\begin{tabular}{lccccccc}
		\toprule
		\textbf{Domain} & $n$ & \textbf{Recursive} & \textbf{Negation} & \textbf{Aggregation} & \textbf{Avg. Strata} & \textbf{Max SCC} & \textbf{Avg. Invented} \\
		\midrule
		Program analysis    & 47 & $42.6\%$ & 10 & \phantom{0}6 & 1.38 & 6 & 1.55 \\
		Graph analytics     & 33 & $75.8\%$ & \phantom{0}9 & \textbf{11} & \textbf{1.82} & \textbf{9} & \textbf{1.67} \\
		Formal reasoning    & 29 & \textbf{79.3\%} & 10 & \phantom{0}4 & 1.62 & 3 & 1.59 \\
		Knowledge discovery & 27 & $37.0\%$ & \phantom{0}5 & \phantom{0}1 & 1.30 & 7 & 0.63 \\
		\bottomrule
	\end{tabular}
	}
\end{table}

\section{Metric Definitions}
\label{app:metrics}

This appendix gives the full form of the metrics of \S\ref{sec:compute}. Let $\mathcal{T}$ denote the evaluated task set, and let $\hat{P}_i$ be the Datalog program generated for task $i \in \mathcal{T}$. We define $C_i=1$ if $\hat{P}_i$ passes compilation without errors and $C_i=0$ otherwise. For executable programs, let $\mathrm{Exec}(\hat{P}_i, x_{ij})$ be the output produced on the $j$-th input variant of task $i$, and let $y_{ij}$ be the corresponding ground-truth output. Generated programs are compiled and executed with Souffl{\'e}. A task is correct only if the generated program compiles and matches the expected output on every evaluation variant. A program that compiles but exceeds the $30$-second evaluation limit on a variant has $C_i=1$; that variant produces no output, so $M_i=0$, its expected tuples count as missed, and its negatives are left out of negative exclusion.

\begin{equation}
	M_i = C_i \cdot \mathbb{I}\left[
	\forall j \in \{1,\ldots,m_i\},\
	\mathrm{Exec}(\hat{P}_i, x_{ij}) = y_{ij}
	\right],
\end{equation}
where $m_i$ is the number of evaluation variants of task $i$ and $\mathbb{I}[\cdot]$ is the indicator function.
\emph{Compile Pass} (CP) and \emph{Exact Match} (EX) are then:
\begin{equation}
	\mathrm{CP}(\mathcal{T}) =
	\frac{1}{|\mathcal{T}|}\sum_{i \in \mathcal{T}} C_i,
	\qquad
	\mathrm{EX}(\mathcal{T}) =
	\frac{1}{|\mathcal{T}|}\sum_{i \in \mathcal{T}} M_i.
\end{equation}
With repeated runs, $C_{i,r}$ and $M_{i,r}$ denote the outcomes of task $i$ in
run $r$, and $\mathrm{CP}_r$ and $\mathrm{EX}_r$ are computed from them as above.
The reported CP and EX are the means over runs, and the tuple-level metrics and
negative exclusion below are likewise computed per run and then averaged.
Every figure reported in the paper uses this pair unless stated otherwise.

\smallskip
\noindent\textbf{Partial credit over derived facts.}
EX is deliberately strict: a program that derives every intended tuple but one scores the same as a program that derives nothing. That is the right primary metric for a synthesis task---a Datalog program that is almost right is still wrong---but it discards information about \emph{how} wrong a failing program is, and with it the ability to distinguish a near-miss from a program that misunderstood the task entirely. We therefore also define tuple-level precision, recall, and $F_1$. Pooling the derived and expected tuples of task $i$ over all of its variants gives
\begin{equation}
\begin{split}
	\mathrm{TP}_i &= \sum_{j} \left| \mathrm{Exec}(\hat{P}_i, x_{ij}) \cap y_{ij} \right|,
	\quad
	\mathrm{FP}_i = \sum_{j} \left| \mathrm{Exec}(\hat{P}_i, x_{ij}) \setminus y_{ij} \right|, \\
	\mathrm{FN}_i &= \sum_{j} \left| y_{ij} \setminus \mathrm{Exec}(\hat{P}_i, x_{ij}) \right|,
\end{split}
\end{equation}
from which precision, recall, and $F_1$ follow in the usual way; a program that fails to compile scores zero on all three. Tuples are compared within the same variant and the same output relation, so two relations of equal arity cannot collide. When a program derives nothing, precision is taken as zero; the pooled expected set is never empty, since every task has at least one expected tuple across its variants. Benchmark-level figures are macro-averages over tasks, so each task counts once regardless of how many tuples it derives, which matters because output sizes are heavy-tailed (\S\ref{sec:stats}). These appear only in this appendix; every comparison in the main text is on CP and EX. Reading them together is what makes them informative: low recall with high precision indicates a missing rule, typically a base case or a recursive step, whereas low precision with high recall indicates an over-general rule, typically a missing join condition or an unconstrained variable.

\smallskip
\noindent\textbf{Negative exclusion.}
Precision and recall reward covering the right tuples, but a synthesis oracle should also ask whether a program avoids the \emph{characteristically wrong} ones. Following the SyGuS formulation of Appendix~\ref{sec:pre}, which specifies both a positive set $O^{+}$ and a negative set $O^{-}$, we define negative exclusion as the fraction of $O^{-}$ that a candidate program does not derive, pooled
per task over its evaluation variants. Let $O^{-}_{ij}$ be the negative set of
variant $j$ of task $i$, with tuples compared within the same output relation, and
let $\mathcal{T}^{-}_{C} = \{i : \sum_j |O^{-}_{ij}| > 0,\ C_i = 1\}$ be the tasks
that carry negatives and whose candidate compiles:
\begin{equation}
	\mathrm{NE}(\mathcal{T}) = \frac{1}{|\mathcal{T}^{-}_{C}|}\sum_{i \in \mathcal{T}^{-}_{C}} \left( 1 - \frac{\sum_{j} \left| \mathrm{Exec}(\hat{P}_i, x_{ij}) \cap O^{-}_{ij} \right|}{\sum_{j} |O^{-}_{ij}|} \right).
\end{equation}
A candidate that does not compile derives nothing, and would trivially exclude
every negative; it is therefore left out of the average rather than scored as
perfect, which makes \textrm{NE} a measure conditional on compiling---one more
reason it is not used for the comparisons in \S\ref{sec:result}. Restricting
the average to tasks with negatives matters as well: $119$ of the $136$ tasks
carry them, $4{,}867$ tuples in all.
Scoring the remaining $17$ as either perfect or zero exclusion would report a
property of the mining procedure as though it were a property of the candidate.
The negatives are \emph{mined rather than invented}: they are the tuples that semantic mutants of the reference program derive and the reference does not (\S\ref{sec:validation}). A candidate that avoids them has avoided the failure modes this task actually invites, rather than a set of tuples perturbed at random---and because they come from the same mutant pool that measures the oracle, both analyses share one definition of a plausible error instead of each adopting its own.

\section{Oracle Strength in Detail}
\label{app:audit}

Section~\ref{sec:validation} compresses the checks and the mutation analysis that
bound what the execution oracle can distinguish. This appendix gives them in
full, together with the current status of surviving mutants.

\subsection{Checks and Mutation Analysis}
\label{app:validation-detail}

\smallskip
\noindent\textbf{Mechanical checks.}
Eleven invariants run in continuous integration over the released files, each
asserting a condition the benchmark could violate silently---that a prompt's
schema matches the reference's declaration, that mined negative sets stay
synchronized after an edit, that single-instance parameters are not scaled like
data. All pass; two further checks report rather than fail, since their findings need a
judgement each time.

\smallskip
\noindent\textbf{Oracle discrimination.}
\label{sec:oracle-strength}
Grading a synthesized program by executing it on finitely many inputs is not a
proof of equivalence: a wrong program can agree with the reference on every
variant we test. This limitation is shared with the text-to-SQL literature and
cannot be removed, but it can be \emph{measured}, and we report that
measurement.

We adopt mutation analysis from software testing. For each reference program we
generate mutants under three operators intended to perturb meaning---deleting a rule,
deleting a body atom, and mis-binding a join variable so a head-shared variable
takes the wrong column. The third is included because a mis-bound join is a
characteristic semantic error in relational programs: it yields a program that
compiles and runs but computes a different relation, which is exactly the kind of
error an execution oracle must catch. A mutant is \emph{killed} if its output differs from the
reference on at least one variant. A surviving mutant may be equivalent to the reference. Automatic checks use
variable renaming or $\theta$-subsumption; an unresolved survivor remains in the
denominator rather than being discounted as equivalent.

Over the $136$ reference programs the three operators generate $1{,}948$
mutants---the complete set they admit, not a sample---of which $1{,}521$ compile,
and the evaluation variants distinguish $1{,}440$ of these, a raw mutation score of $94.7\%$.
Of the $70$ survivors, $13$ are equivalent to the reference by an automatic check;
the rest have not been checked one by one, so the score is a lower bound. Two limits bound
the claim further: three operators do not span the error space, and a high score
shows the oracle separates the reference from \emph{these} mutants, not from an
arbitrary wrong program.

The first evaluation instance distinguishes $1{,}246$ of the $1{,}521$ compilable
mutants ($81.9\%$), and additional instances distinguish another $194$, adding
$12.8$ percentage points. The benchmark contains $251$ evaluation instances and
$136$ separate demonstration instances.

\smallskip
\noindent\textbf{Negative coverage.}
The same mutants supply the negative sets defined in \S\ref{sec:compute}.
Only over-deriving mutants contribute negative tuples, so this coverage is
uneven: deleting a rule normally under-derives, whereas deleting a body atom or
rebinding a join may over-derive when the rule remains safe and executable.

\subsection{Status of Surviving Mutants}
\label{app:survivors}

Of the $70$ surviving mutants from the original operators, automatic checks
classify $13$ as equivalent; the remaining $57$ have not been checked one by one.
None is removed from the reported denominator. We therefore report the
unadjusted $94.7\%$ discrimination rate as a lower bound.

\subsection{Additional Mutation Operators}
\label{app:new-operators}

The three operators above also guided the choice of evaluation variants, so
their score measures the variants against the mutants that selected them. We
therefore apply five further operators, each changing one site of one rule:
\begin{itemize}
	\item \emph{negation removal} drops the negation of one negated atom;
	\item \emph{aggregate replacement} swaps the aggregate function
	(\texttt{min} $\leftrightarrow$ \texttt{max}, \texttt{sum} $\rightarrow$
	\texttt{max}, \texttt{mean} $\rightarrow$ \texttt{sum});
	\item \emph{grouping change} renames, inside an aggregate, one variable shared
	with the rest of the rule, so the aggregate no longer groups on it;
	\item \emph{comparison bound} moves a boundary ($<$ $\leftrightarrow$ $\le$,
	$>$ $\leftrightarrow$ $\ge$) or turns $\ne$ into $=$;
	\item \emph{constant change} replaces an integer $n$ by $n+1$ or $n-1$;
		for a string literal, it selects another literal from the same program.
\end{itemize}
A mutant that Souffl{\'e} rejects is counted separately and is neither
distinguished nor surviving. Both groups are evaluated on the frozen set of $251$ variants. Table~\ref{tab:mutation-operators} reports each
operator. Together the five operators generate $800$ mutants that compile, of
which the evaluation variants distinguish $693$ ($86.6\%$). We have not checked the
$91$ survivors for equivalence, so this is a lower bound.

\begin{table}[htbp]
	\centering
	\small
	\caption{Mutation analysis by operator. Programs counts the reference programs
	with at least one applicable site; Rejected counts mutants that Souffl{\'e}
	does not compile. Compiled includes mutants that fail only at run time, which
	count as neither distinguished nor surviving ($11$ and $16$).}
	\label{tab:mutation-operators}
	\setlength{\tabcolsep}{3.5pt}
	\begin{tabular}{lcccccc}
		\toprule
		Operator & Programs & Mutants & Rejected & Compiled & Distinguished & Survivors \\
		\midrule
		\multicolumn{7}{l}{\emph{Original operators}} \\
		Rule or body-atom deletion & 136 & 1{,}147 & 329 & 818 & 783 & 31 \\
		Join rebinding & 129 & 801 & 98 & 703 & 657 & 39 \\
		Subtotal & 136 & 1{,}948 & 427 & 1{,}521 & 1{,}440 & 70 \\
		\midrule
		\multicolumn{7}{l}{\emph{Additional operators}} \\
		Negation removal & 36 & 72 & 0 & 72 & 69 & 2 \\
		Aggregate replacement & 15 & 26 & 0 & 26 & 25 & 1 \\
		Grouping change & 18 & 28 & 6 & 22 & 21 & 1 \\
		Constant change & 50 & 556 & 0 & 556 & 501 & 43 \\
		Comparison bound & 37 & 124 & 0 & 124 & 77 & 44 \\
		Subtotal & 77 & 806 & 6 & 800 & 693 & 91 \\
		\bottomrule
	\end{tabular}
\end{table}

\subsection{Independently Generated Test Inputs}
\label{app:fresh-tests}

We fixed the scope before generating any input: the $58$ tasks that had a single
evaluation variant at that point. A generic generator samples shapes and values
from each task's pools; a structured generator builds well-formed trees, grids,
and related structures for $12$ tasks; hand-specified comparison-boundary cases
cover three tasks. No generator reads a candidate program.

Admission checks enforce declared arities, one tuple per parameter relation,
and input constraints encoded from the task statement. After normalization, an
input must duplicate no demonstration, evaluation, or admitted input, and its
reference output must be non-empty within the time limit. We uniformly sample
at most ten admitted generic or structured inputs per task with a fixed seed,
without using mutant or candidate scores. The structured generator does not
produce nested-function cases.

\begin{samepage}
Across three runs of each configuration, we retested all $3{,}434$ accepted
programs on these $58$ tasks: $3{,}318$ pass the additional inputs and $116$ are
rejected. Inspection identifies $6$ wrong programs, $65$ cases left undefined by
the task statement, and $45$ cases with two readings. Encoded input constraints
do not resolve these conventions.\par
\end{samepage}

The six wrong programs all come from DeepSeek V4 Flash without thinking.
Counting them as failures lowers a configuration's mean EX by at most $1.0$ point,
with other models unchanged. Counting all $116$ rejections as failures lowers
it by at most $2.5$ points. Agent and example gains, and the recursion and
auxiliary-predicate gaps, retain their direction. The descriptions-versus-signatures
difference remains inconclusive, with its interval including zero.

\begin{table}[htbp]
\centering\small
\caption{Additional test inputs and accepted programs retested across three runs
of each configuration. A program can appear under more than one generator;
the All row counts each once. The $116$ rejections comprise $6$ wrong programs,
$65$ undefined cases, and $45$ cases with two readings.}
\label{tab:fresh-tests}
\begin{tabular}{lrrrr}
\toprule
Generator & Tasks & Inputs & Programs tested & Rejected \\
\midrule
Generic & 46 & 373 & 2{,}752 & 90 \\
Structured & 12 & 120 & 682 & 26 \\
Boundary & 3 & 6 & 186 & 0 \\
\midrule
All & 58 & 499 & 3{,}434 & 116 \\
\bottomrule
\end{tabular}
\end{table}

\section{Code, Environment, and Model Provenance}
\label{app:artifact}

This appendix records what is needed to rebuild the evaluation from a clean
checkout, and states one limitation of our model access that affects how
exactly our numbers can be reproduced.

\subsection{Building the Symbolic Baselines}
\label{app:baselines}

The three symbolic synthesizers are included as pinned Git submodules rather
than vendored copies, so the third-party sources stay attributable to their
upstream repositories at a fixed commit. Building them on a current toolchain
requires two source-level fixes, which we ship as committed patches rather than
as edits to the submodules: one pins the sbt version used to assemble the EGS
fat jar, and one removes a duplicate \texttt{AC\_CONFIG\_MACRO\_DIR} declaration
that modern \texttt{autoconf} rejects when configuring the Souffl{\'e} build
that ProSynth depends on.

A single setup script performs the whole sequence---initialize the submodules,
apply each patch idempotently, then build each tool---so preparing the baselines
from a fresh clone is one command. Patch application is checked in both
directions before it runs, so re-running the script on an already-prepared tree
is a no-op rather than an error. Building the three tools requires no container
runtime; the host needs a JDK 11 or newer together with sbt for
EGS (its jar uses \texttt{String.strip}), an SMT solver and the patched
Souffl{\'e} for ProSynth, and Python with Souffl{\'e} for GenSynth. The script
skips, with a diagnostic, any tool whose prerequisites are absent, so a partial
environment still yields a usable subset rather than a failed build.

\subsection{Execution Environment}

Generated and reference programs are compiled and executed with
Souffl{\'e}~2.5 under Python~3.11.2. Every execution---one program on one input
variant---is subject to a wall-clock timeout of $30$ seconds, enforced by the
harness and overridable through an environment variable for machines
substantially slower than ours; an execution that exceeds it is recorded as a
failure for that variant rather than aborting the run. All evaluation is
CPU-only: no model is trained or fine-tuned, and the reported experiments
involve inference against hosted APIs plus local Souffl{\'e} execution.

\subsection{Model Access and Version Provenance}
\label{app:provenance}

We record provenance rather than assume it. For every
direct-prompting call, the harness stores the model identifier the service
returned alongside the generation, and each run reports the distribution of
served identifiers it observed, so a divergence between the requested name and
the served build, or a change in that mapping between runs, is visible in the
record. For DeepSeek, both Flash arms send the same API model name,
\texttt{deepseek-v4-flash}, and differ only in an explicit \texttt{thinking}
request field; the arms are kept under distinct experiment names so their result
directories cannot overwrite each other. Across the $3{,}264$ calls of the two
arms, the returned identifier is \texttt{deepseek-flash} without exception,
with reasoning tokens present on all $1{,}632$ thinking calls and on none of the
$1{,}632$ non-thinking ones, so the ablation is a mode switch on one served model.
For V4~Pro, requests use \texttt{deepseek-v4-pro} at an explicit low reasoning effort,
and the service returns \texttt{deepseek-v4-pro} on every call. For the other three
models the returned identifier matches the requested one on every call that
returned. Gemini was evaluated through the official API. DeepSeek~V4~Flash also has
an open-weight \texttt{0731} revision, but our measurements use the hosted API
rather than a local deployment.

The coding agents are the exception. They are driven through their vendors'
command-line interfaces, which report no server-side model identifier, so for
that pipeline we record only the requested model. The agent results in
\S\ref{sec:agent} are therefore attributable to a provider and a requested
model name, not to a verified build.

Two limits apply even where provenance is recorded. Reproduction is exact only
insofar as the same served build is still reachable; a later run under the same
model name may resolve to a different checkpoint, and while our records make that
detectable, they cannot make the old checkpoint available again. And providers
differ in how precisely they identify a build, so a recorded identifier is only
as specific as its provider chooses to be. These served-model identification
limits do not apply to the locally executed symbolic tools or the Souffl{\'e}
evaluation runtime.

\section{What the Failures Look Like}
\label{app:qualitative}

Aggregate error rates say how often programs fail and under which label; they do
not show what a failing program says. This appendix retains five illustrative outputs from the original
evaluation, selected by a fixed rule: one case per error class---an undeclared auxiliary predicate, a
misused library predicate, an aggregate without its grouping atom, an
under-derivation, and an over-derivation---and within each class a case whose
reference and generated programs both fit in a few lines, so the two can be read
side by side. Three of the six models appear. We did not select for severity: two
of the five are failures of the strongest model in the study.

\smallskip
\noindent\textbf{The auxiliary predicate that was never declared.}
\texttt{Bigrams} asks for adjacent word pairs in a sequence. The reference reads
the position and adds one to it. Claude~Opus~5 instead invents an
\texttt{Adjacent} relation to mean ``no position lies strictly between,'' and
never declares it, so the program does not compile.

\begin{lstlisting}
% reference
Bigram(w, w1) :- Words(i, w), Words(i+1, w1).

% Claude Opus 5, signature-only -- Error: Undefined relation Adjacent
Adjacent(p1, p2) :- Words(p1, _), Words(p2, _), p1 < p2,
                    !Words(p3, _), p1 < p3, p3 < p2.
Bigram(w1, w2) :- Adjacent(p1, p2), Words(p1, w1), Words(p2, w2).
\end{lstlisting}

\noindent Two things are worth separating. The decomposition is not unreasonable:
naming adjacency is how one would explain the task aloud. What fails is
compound rather than superficial. \texttt{Adjacent} lacks a \texttt{.decl}, but
the negation is also unsafe because \texttt{p3} is never bound by a positive
literal. Adding the declaration alone therefore does not make this program
compile, much less make it exact. The example illustrates why declaration-only
repair is reported as a bounded diagnostic rather than treated as semantic
repair (\S\ref{sec:failure}).

\smallskip
\noindent\textbf{The library predicate called backward.}
\texttt{GetImportSpringFrameworkFileJava} asks which Java files import Spring.
GPT~5.6~Sol's program is the reference program---same joins, same variables, same
output---with one built-in's arguments transposed.

\begin{lstlisting}
% reference
ImportStatement(name, file_path) :-
    JavaImport(_, name, file_id), contains("springframework", name),
    JavaFile(file_id, file_path).

% GPT 5.6 Sol, signature-only -- compiles, derives nothing
ImportStatement(name, file_path) :-
    JavaImport(_, name, file_id), JavaFile(file_id, file_path),
    contains(name, "springframework").
\end{lstlisting}

\noindent Souffl{\'e} reads \texttt{contains(sub, str)} as ``\texttt{sub} occurs in
\texttt{str}.'' Transposed, the rule asks whether each import name occurs inside
the literal \texttt{"springframework"}, which nothing satisfies: the program
compiles, runs, and returns the empty relation ($\mathrm{tp}=0$,
$\mathrm{fn}=2$). This is not one model's slip. Across the three tasks whose
reference calls \texttt{match} or \texttt{contains} with one literal and one
variable, $37$ of the $132$ generated calls transpose the arguments, from five of
the six models and in all four prompt settings. The logic is right and the calling
convention is wrong, and no amount of reasoning about the relation recovers it,
because the fact needed is about a standard library rather than about the task.

\smallskip
\noindent\textbf{The aggregate without its grouping literal.}
\texttt{ShortestEdges} asks for each node's shortest outgoing edge. The reference
grounds the node in an outer atom and then minimizes within it.

\begin{lstlisting}
% reference
ShortestEdge(x, l) :- Edge(x, _, _), l = min z : Edge(x, _, z).

% DeepSeek V4 Pro, one-shot description -- tp = 2, fn = 1601
ShortestEdge(node, min weight : Edge(node, _, weight)).
\end{lstlisting}

\noindent Dropping \texttt{Edge(x, \_, \_)} does not drop a redundant atom; it
removes the only thing that makes the minimum per-node, collapsing a grouped
aggregate into a global one. The program derives $2$ tuples where $1{,}603$ are
expected. The generated line reads as a faithful transcription of ``the minimum
weight of the edges out of \texttt{node}''---the error is invisible in English and
total in Datalog.

\smallskip
\noindent\textbf{Symmetry assumed, and symmetry omitted.}
The last two cases are the same confusion in opposite directions, which is why we
print them together. \texttt{Traffic} asks which roads crash: a road with traffic
that can collide with another road that also has traffic. \texttt{Intersect} is
not symmetric in the data, so the reference needs both orientations.

\begin{lstlisting}
% reference
Crashes(x) :- HasTraffic(x), Crashable(x, y), HasTraffic(y).
Crashes(x) :- HasTraffic(x), Crashable(y, x), HasTraffic(y).

% GPT 5.6 Sol, signature-only -- tp = 7, fp = 0, fn = 2
Crashes(road_a) :- HasTraffic(road_a), Crashable(road_a, road_b),
                   HasTraffic(road_b).
\end{lstlisting}

\noindent The model wrote \texttt{Crashable} exactly right and then supplied one
of the two \texttt{Crashes} rules, missing every road that appears only as the
second argument of a colliding pair. It derives too little; across the reported runs, the task fails in
$62$ of the $72$ Direct evaluations.
\texttt{RevSameGen} then fails the other way. Its reference threads the recursive
atom between the ancestor and descendant links; Claude~Opus~5 chains it
left-to-right instead, adds a fourth rule for the reverse traversal, and closes
the base case symmetrically.

\begin{lstlisting}
% reference
RevSameGen(x, y) :- Flat(x, y).
RevSameGen(x, y) :- Up(x, a), RevSameGen(b, a), Down(b, y).

% Claude Opus 5, signature-only -- tp = 3203, fp = 3200, fn = 7
RevSameGen(x, y) :- Flat(x, y).
RevSameGen(x, y) :- Flat(y, x).
RevSameGen(x, y) :- RevSameGen(x, z), Up(z, w), Down(w, y).
RevSameGen(x, y) :- RevSameGen(x, z), Down(z, w), Up(w, y).
\end{lstlisting}

\noindent Recall is nearly perfect and precision is about a half: the program
derives roughly twice the intended relation. Read against \texttt{Traffic}, the
pair says something the aggregate counts cannot. Neither model derived symmetry
from the specification. One assumed a relation named ``same generation'' must be
symmetric and closed it; the other assumed a collision relation was already
symmetric and did not. The specification settles both cases, yet both models
answered from the name.

\section{Compile-Failure Taxonomy}
\label{app:error-taxonomy}

Section~\ref{sec:failure} groups the Direct compile failures by cause;
Table~\ref{tab:error-taxonomy} gives the classes behind each group. Every failure
is traced to the statement Souffl{\'e} rejects in the program that was actually
compiled, and assigned by fixed rules on that statement and the error message. A
type-inference failure counts toward predicate invention only when the rejected
rule defines a relation outside the schema; otherwise it counts toward typing.
Programs that compile and then exceed the evaluation timeout are not compile
failures and are excluded. We checked a sample of each class by hand.

\begin{table}[h]
	\centering
	\caption{Causes of the $2{,}665$ Direct compile failures, pooled over the $24$ cells and the \nruns{} runs.}
	\label{tab:error-taxonomy}
	\small
	\begin{tabular}{p{3.6cm}lrr}
		\toprule
		Group & Class & Count & Share (\%) \\
		\midrule
		Predicate invention left incomplete & Undeclared auxiliary predicate & 986 & 37.0 \\
		 & Untypable rule defining one & 445 & 16.7 \\
		\midrule
		Souffl{\'e} dialect & Aggregate computed in a rule head & 249 & 9.3 \\
		 & Malformed body aggregate & 87 & 3.3 \\
		 & Reserved word used as a name & 115 & 4.3 \\
		 & Operator from C or SQL & 11 & 0.4 \\
		 & \texttt{not} instead of \texttt{!} & 139 & 5.2 \\
		 & Other syntax & 87 & 3.3 \\
		\midrule
		Typing of constants and schema relations & Constant of the wrong type & 322 & 12.1 \\
		 & Untypable rule defining a schema relation & 20 & 0.8 \\
		\midrule
		Safety and stratification & Ungrounded variable & 98 & 3.7 \\
		 & Negation that cannot be stratified & 56 & 2.1 \\
		\midrule
		Response not a clean program & Response cut off mid-statement & 13 & 0.5 \\
		\midrule
		Other & Unclassified & 37 & 1.4 \\
		\bottomrule
	\end{tabular}
\end{table}

\section{Structural Analysis of Failures}
\label{sec:analysis}

Section~\ref{sec:failure} states the two headline gaps; this appendix gives the
measurements behind them.

Two structural perspectives take the failure analysis past aggregate exact match:
recursion and predicate invention. Both are axes on which Datalog differs from the
query languages code-generation benchmarks usually cover, and both are properties
\bench was built to expose.

\smallskip
\noindent \textbf{Recursion is where models fail, but not because they miss it.}
Of the $136$ reference programs, $78$ contain a cycle in their predicate
dependency graph and $58$ do not. Under signature-only zero-shot prompts, exact
match on recursive tasks ranges from $23.5\%$ to $51.3\%$ across the six models,
against $67.8\%$--$85.1\%$ on non-recursive tasks. The corresponding gaps are
$31.3$--$46.6$ points. The generated dependency graphs that can be parsed remain
highly recursion-aware (the model usually creates a recursive cycle when the
reference has one), so the outcome gap is not explained simply by omitting
recursion. The harder problem is constructing the right base case, inductive
step, and variable bindings. The agent gains are consistent with this reading:
most gains not explained by declaration repair occur on recursive tasks
(Appendix~\ref{app:agent-detail}).

\smallskip
\noindent \textbf{Auxiliary-predicate structure marks a second gap.}
Counting every relation defined in a rule head that is neither input nor output,
$59$ of the $136$ reference programs ($43.4\%$) introduce at least one auxiliary
predicate, $189$ in all, at a mean of $1.4$ per task; $23$ introduce three or
more and the largest uses $19$.
This label describes the reference program; it does not assert that every
correct solution must introduce an auxiliary predicate.

Exact match on the $59$ tasks whose reference introduces an auxiliary predicate
ranges from $20.9\%$ to $53.1\%$, against $60.2\%$--$75.3\%$ on the $77$
whose references use none. Measuring how many predicates candidates invent requires care because
structural parsing excludes malformed programs, and the excluded programs are
disproportionately those that omit declarations. Overlaying the repaired
declarations raises parser coverage, and with it the measured invention rate of
the models that omit declarations most often, without changing any rule. We therefore use the reference-side binary
property for outcome slices and do not interpret candidate invention counts as a
causal measure.

The two gaps must not be added, because the two sets of tasks largely coincide:
$47$ of the $59$ tasks whose references use an auxiliary predicate are also recursive, leaving
$31$ recursive tasks whose references use none, $12$ non-recursive tasks whose references use one or more, and $46$
tasks whose references have neither property. Each gap nonetheless survives with the other property held
fixed. Jointly, exact match falls from $73.9\%$--$89.1\%$ on the $46$ tasks with
neither property to $17.7\%$--$48.9\%$ on the $47$ with both. The split is observational---tasks with
these properties also differ in size and domain---so the gaps locate where
models fail rather than isolate a cause. The paragraphs below control for
program complexity and examine the structure of accepted programs. Difficulty on \bench is concentrated in
two overlapping structural demands:
expressing a fixed point correctly, and constructing useful decompositions through auxiliary predicates.

\smallskip
\noindent\textbf{The gaps with program complexity held fixed.}
Recursive tasks and tasks whose references use auxiliary predicates also have
larger references, so we re-estimate both gaps with complexity controlled. The
covariates come from the reference program and the specification and exclude the
quantities that define the two labels: the number of strata, the mean number of
body atoms per rule, the largest arity, the number of input relations, and the
length of the question in words. Rule count enters only as a sensitivity check,
because it partly follows from the labels: a recursive definition needs a base
rule and a step rule, and an auxiliary predicate needs rules of its own.

We use two estimators. The \emph{stratified} estimate sorts tasks by the mean
percentile rank of the covariates, cuts them into three equal-count strata,
splits each by the other label, takes the exact-match difference within each
cell, and pools the cells with Mantel--Haenszel weights. The \emph{regression}
estimate fits a logistic model of per-task success on both labels, the
standardized covariates, and domain indicators, and reports the odds ratio and
the average marginal effect in exact-match points. Intervals are bootstrap
intervals over tasks (Appendix~\ref{app:reproducibility}).
Table~\ref{tab:controlled-gaps} reports both gaps. Both remain for the Direct cells
with complexity controlled, with $95\%$ intervals that exclude zero. For each agent
the stratified recursion gap also excludes zero, whereas the stratified
auxiliary-predicate gap does not.

\smallskip
\noindent\textbf{Exact match by program-complexity tier.}
Using the descriptive tiers of Table~\ref{tab:difficulty}, Direct EX falls from
$73.5\%$ to $32.7\%$ from easiest to hardest; Codex falls from $94.2\%$ to $64.4\%$
and CC from $94.9\%$ to $73.3\%$. These tiers include recursive-component size;
the estimates above instead control for covariates that do not define the labels.

\smallskip
\noindent\textbf{Do accepted programs use the labeled construct?}
We analyze the structure of all $6{,}277$ programs accepted by the held-out
evaluation across the reported runs. Table~\ref{tab:correct-structure} relates
the reference labels to the accepted programs. On recursive tasks $98.3\%$ of the accepted
programs are recursive, against $5.7\%$ on the other tasks; on auxiliary-predicate
tasks $63.5\%$ introduce a predicate, against $5.4\%$ on the other tasks. Eight
recursive-labeled tasks have an accepted program without recursion, and $25$
auxiliary-labeled tasks have one without auxiliary predicates.

\begin{table}[htbp]
	\centering
	\small
	\caption{Recursion and auxiliary-predicate gaps in exact match (points), raw and
	with program complexity controlled, over \nruns{} runs. Direct pools the $24$ Direct cells. AME is
	the average marginal effect; brackets give $95\%$ intervals over tasks.}
	\label{tab:controlled-gaps}
	\setlength{\tabcolsep}{4pt}
	\resizebox{\textwidth}{!}{%
	\begin{tabular}{llccccc}
		\toprule
		& & & \multicolumn{2}{c}{Stratified gap} & \multicolumn{2}{c}{Regression} \\
		\cmidrule(lr){4-5}\cmidrule(lr){6-7}
		Label & Setting & Raw gap & Primary & With rule count & Odds ratio & AME \\
		\midrule
		\multirow{3}{*}{Recursion} & Direct & $-35.6$ & $-27.8$ {\scriptsize[$-38.9$, $-17.1$]} & $-27.6$ & $0.29$ & $-25.6$ {\scriptsize[$-37.3$, $-14.6$]} \\
		 & Codex & $-18.5$ & $-14.1$ {\scriptsize[$-26.4$, $-2.0$]} & $-13.3$ & $0.38$ & $-9.5$ {\scriptsize[$-21.4$, $2.7$]} \\
		 & CC & $-16.1$ & $-12.5$ {\scriptsize[$-21.9$, $-3.5$]} & $-12.9$ & $0.26$ & $-10.9$ {\scriptsize[$-21.2$, $-1.7$]} \\
		\midrule
		\multirow{3}{*}{Auxiliary predicate} & Direct & $-34.4$ & $-18.1$ {\scriptsize[$-29.9$, $-5.4$]} & $-16.9$ & $0.49$ & $-14.6$ {\scriptsize[$-27.7$, $-0.8$]} \\
		 & Codex & $-22.0$ & $-10.7$ {\scriptsize[$-24.0$, $3.1$]} & $-10.8$ & $0.33$ & $-11.4$ {\scriptsize[$-24.5$, $0.7$]} \\
		 & CC & $-15.5$ & $-4.1$ {\scriptsize[$-14.4$, $6.5$]} & $-6.2$ & $0.47$ & $-6.8$ {\scriptsize[$-18.0$, $3.1$]} \\
		\bottomrule
	\end{tabular}}
\end{table}

\begin{table}[htbp]
	\centering
	\small
	\caption{Structure of accepted programs against the reference labels, over all
	runs. With construct counts the accepted programs that are themselves recursive
	(upper rows) or introduce an auxiliary predicate (lower rows). Differing tasks
	counts tasks with at least one accepted program whose structure differs from the
	reference label.}
	\label{tab:correct-structure}
	\setlength{\tabcolsep}{4pt}
	\begin{tabular}{lcccc}
		\toprule
		Reference label & Tasks & Accepted programs & With construct & Differing tasks \\
		\midrule
		Recursive & 78 & 2{,}714 & 2{,}667 & 8 \\
		Non-recursive & 58 & 3{,}563 & 204 & 11 \\
		\midrule
		Auxiliary predicate & 59 & 1{,}859 & 1{,}180 & 25 \\
		No auxiliary predicate & 77 & 4{,}418 & 239 & 31 \\
		\bottomrule
	\end{tabular}
\end{table}

\section{Per-Domain Model Performance}
\label{app:per-domain}

Table~\ref{tbl:base} reports overall Compile Pass and Exact Match;
Table~\ref{tab:per-domain} gives the same runs split by reasoning domain, abbreviated as KD (knowledge
discovery), GA (graph analytics), PA (program analysis) and FR (formal reasoning).
The breakdown shows that no single domain ordering holds for every model. Knowledge
discovery is the easiest domain in eleven of the twelve model--schema pairs. Formal reasoning is the
weakest in nine of the twelve model--schema pairs; in the other three, graph analytics or
program analysis is lower. This is why the main analysis uses explicit structural properties
rather than treating domain labels as a difficulty scale.

\begin{table}[t]
	\centering
	\caption{Zero-shot performance by domain under both schemas, as means over \nruns{} runs.}
	\label{tab:per-domain}
	\resizebox{0.98\textwidth}{!}
	{
\begin{tabular}{llccccc|ccccc}
			\toprule
			\multirow{2}{*}{Model} & \multirow{2}{*}{Schema} &
			\multicolumn{5}{c}{Compile Pass} &
			\multicolumn{5}{c}{Exact Match} \\
			\cmidrule(lr){3-7}\cmidrule(lr){8-12}
			& & KD (\%) & GA (\%) & PA (\%) & FR (\%) & \textbf{Overall}
			  & KD (\%) & GA (\%) & PA (\%) & FR (\%) & \textbf{Overall} \\
			\midrule
\multirow{2}{*}{GPT~5.6~Sol} & Signature & 91.4 & 61.6 & 74.5 & 72.4 & 74.3 & 70.4 & 49.5 & 50.4 & 46.0 & 53.2 \\
			 & Description & 86.4 & 64.6 & 77.3 & 74.7 & 75.5 & 75.3 & 50.5 & 50.4 & 50.6 & 55.4 \\
			\midrule
			\multirow{2}{*}{Claude Opus~5} & Signature & 88.9 & 85.9 & 85.1 & 82.8 & 85.5 & 67.9 & 66.7 & 65.2 & 56.3 & 64.2 \\
			 & Description & 91.4 & 79.8 & 84.4 & 80.5 & 83.8 & 72.8 & 57.6 & 63.8 & 58.6 & 63.0 \\
			\midrule
			\multirow{2}{*}{Gemini~3.8~Flash} & Signature & 91.4 & 90.9 & 82.3 & 88.5 & 87.5 & 70.4 & 63.6 & 71.6 & 54.0 & 65.7 \\
			 & Description & 92.6 & 85.9 & 83.7 & 85.1 & 86.3 & 75.3 & 67.7 & 67.4 & 58.6 & 67.2 \\
			\midrule
			\multirow{2}{*}{DeepSeek~V4~Pro} & Signature & 81.5 & 56.6 & 63.8 & 49.4 & 62.5 & 64.2 & 41.4 & 45.4 & 33.3 & 45.6 \\
			 & Description & 80.2 & 57.6 & 61.0 & 51.7 & 62.0 & 70.4 & 47.5 & 44.7 & 37.9 & 49.0 \\
			\midrule
			\multirow{2}{*}{DeepSeek~V4~FT} & Signature & 81.5 & 54.5 & 53.9 & 56.3 & 60.0 & 65.4 & 40.4 & 44.7 & 33.3 & 45.3 \\
			 & Description & 82.7 & 50.5 & 56.0 & 59.8 & 60.8 & 67.9 & 38.4 & 47.5 & 42.5 & 48.3 \\
			\midrule
			\multirow{2}{*}{DeepSeek~V4~FNT} & Signature & 84.0 & 69.7 & 73.8 & 41.4 & 67.9 & 59.3 & 40.4 & 48.9 & 23.0 & 43.4 \\
			 & Description & 79.0 & 65.7 & 81.6 & 52.9 & 71.1 & 58.0 & 41.4 & 51.8 & 32.2 & 46.3 \\
			\bottomrule
		\end{tabular}
	}
\end{table}

\section{Symbolic Synthesis from Examples}
\label{app:symbolic}

We also run three non-LLM synthesizers spanning the dominant symbolic strategies:
GenSynth~\citep{mendelson2021gensynth} (genetic search),
EGS~\citep{thakkar2021example} (example-guided) and
ProSynth~\citep{raghothaman2020provenance} (provenance-guided CEGIS).
They consume examples rather than language, so each receives only the example
component and, as in the tools' original setting where the input--output examples
are the specification, learns from the first evaluation variant of each task; it is
then graded by the same oracle on all evaluation variants. The example a tool learns
from is therefore also scored, and for the $58$ single-variant tasks it is the only
variant, so their EX is an optimistic reference. Because a symbolic tool can return
no program at all, we also report a \emph{synthesis success rate}. The tools emit rules without Souffl{\'e}
declarations, so the composition harness supplies declarations for any predicates
they introduce. Their EX is consequently comparable in spirit to Direct
EX$^\dagger$, not raw Direct EX. Each search has a $300$-second per-task timeout.
The pairing is not asking which approach
is better---the tools never see the language and the models never see the search
space---but what the natural-language interface costs. A task the tools solve and
the models do not have a target reachable from their examples, locating the
difficulty in recovering the program \emph{from the description}; the reverse case
identifies tasks where the question carries information the examples do not.
ProSynth is additionally bounded by body-literal count
(\texttt{-{}-rule\_width}~$=2$), so its results describe that hypothesis space
rather than the tool unconstrained (Appendix~\ref{app:model-details}).

\begin{table}[t]
	\centering
	\caption{Symbolic synthesis from examples, as percentages over the $136$ tasks.
	Run counts are given in the Runs column. Here, $\pm$ denotes the wider half of
	the task-level percentile-bootstrap $95\%$ interval, not a run-level standard
	deviation. Synthesis success means returning a non-empty rule set.}
	\label{tab:symbolic}
	\resizebox{0.94\textwidth}{!}{
		\begin{tabular}{lccccc}
			\toprule
			Tool & Runs & Synthesis success (\%) &
			CP (\%) & EX (\%) & Average $F_1$ \\
			\midrule
			EGS      & 1 & $58.8 \pm 8.1$ & $100.0 \pm 0.0$ & $36.0 \pm 8.1$ & $45.3 \pm 7.9$ \\
			GenSynth & 5 & $34.6 \pm 7.8$ & $100.0 \pm 0.0$ & $25.4 \pm 7.2$ & $30.6 \pm 7.2$ \\
			ProSynth & 1 & $10.3 \pm 5.1$ & $100.0 \pm 0.0$ &  $8.1 \pm 5.1$ &  $9.4 \pm 5.1$ \\
			\bottomrule
		\end{tabular}
	}
\end{table}

\smallskip
\noindent\textbf{Returning a program and solving the task are distinct events.}
Table~\ref{tab:symbolic} separates the two. EGS returns a non-empty rule set for
$80$ tasks and exactly solves $49$; GenSynth returns one for $34.6\%$ of tasks on
average and solves $25.4\%$; ProSynth returns one for $14$ and solves $11$. The
identical CP rates are a composition artifact: a failed search can leave an
empty or incomplete candidate that Souffl{\'e} accepts after interface assembly,
even though it derives the wrong relation. These rows compare specifications rather than systems---the tools see one
instance and no question, the models see the question and schema but not the
search space---so EGS's higher EX within the group does not establish that
example-guided search beats language-model synthesis, and ProSynth's figure
describes its \texttt{-{}-rule\_width}~$=2$ hypothesis space rather than the tool
unconstrained.

\Needspace{12\baselineskip}
\section{Agent Turn-by-Turn Detail}
\label{app:agent-detail}

\noindent\textbf{Turn budget and selection.} Sessions run for up to eight turns,
stopping when a candidate is exact on the demonstration instance. Candidates are
ranked by demonstration EX, tuple-level $F_1$, compilation, then fewest differing
tuples. Four-turn results retain only the first four turns of the same
trajectories---one initial generation and up to three feedback-guided
revisions---and apply the same selection rule. The prompt does not state the budget, and later turns cannot affect
earlier ones. Timeout or empty-response failures are retried once in a fresh
session; none occurred after turn four in these runs.

Table~\ref{tab:agent-budget} compares both budgets. Only $7.0$ Codex and $3.7$ CC
sessions per run continue beyond four turns. The extra turns raise mean EX by
$1.5$ points for Codex and $0.0$ for CC. Two CC sessions in one run fail after
retry and count as zero in both views.

\begin{table}[htbp]
\centering\small
\caption{Agent budget comparison over three runs. EX is mean $\pm$ sample
standard deviation (\%); session counts are means per $136$ tasks.}
\label{tab:agent-budget}
\begin{tabular}{lcc}
\toprule
 & Codex / GPT~5.6~Sol & CC / Claude Opus~5 \\
\midrule
Four-turn EX & $83.1\pm1.3$ & $86.8\pm2.2$ \\
Eight-turn EX & $84.6\pm0.0$ & $86.8\pm2.2$ \\
Sessions using turns beyond four & 7.0 & 3.7 \\
\bottomrule
\end{tabular}
\end{table}

Under four turns, $129.0$ Codex and $131.7$ CC candidates per run are exact on
the demonstration; held-out variants reject $16.7$ and $15.0$. Conversely, $0.7$
and $1.3$ are exact on the evaluation variants but not on the demonstration.
This supports keeping evaluation instances withheld.

\smallskip
\noindent\textbf{Composition of the gain.}
Table~\ref{tab:agent-feedback} splits the tasks each agent solves beyond its
model's direct prompt, and slices exact match by the structure of the reference.
A task counts as solved by declaration repair if synthesizing the missing
\texttt{.decl} lines of the program from the Direct cell, the procedure of
\S\ref{sec:compute}, makes it exact; that repair calls no model and
iterates nothing. A few tasks go the other way, solved by the Direct cell (with repair) but not by the
agent: $0.3$ per run for Codex and $1.3$ for CC. Among the tasks not explained by declaration
repair, recursive ones dominate. As in \S\ref{sec:failure}, the split is
observational: the gaps locate where failures remain, not what causes them.

\begin{table}[h]
	\centering
	\caption{What agent interaction changed with relation signatures and no input--output example, against each
	agent's matched Direct cell. Upper block: tasks the agent solves and the Direct
	cell does not, split by whether declaration repair alone solves them. Lower block: exact match (\%) by reference structure, with each pair's gap
	computed from the counts before percentages are rounded. Means over \nruns{} runs.}
	\label{tab:agent-feedback}
	\small
	\begin{tabular}{lcc}
		\toprule
		& Codex / GPT~5.6~Sol & CC / Claude Opus~5 \\
		\midrule
		Solved by agent, not by Direct & $41.0$ & $32.0$ \\
		\quad already solved by declaration repair & $9.0$ ($22.0\%$) & $0.0$ ($0.0\%$) \\
		\quad not explained by declaration repair & $32.0$ & $32.0$ \\
		\quad\quad of which recursive / auxiliary predicate & $23.7$ / $14.3$ & $23.7$ / $16.3$ \\
		\midrule
		Recursive ($78$): Direct $\rightarrow$ agent & $37.2 \rightarrow 75.2$ & $50.9 \rightarrow 79.9$ \\
		Non-recursive ($58$) & $74.7 \rightarrow 93.7$ & $82.2 \rightarrow 96.0$ \\
		\quad recursion gap & $37.5 \rightarrow 18.5$ & $31.3 \rightarrow 16.1$ \\
		Auxiliary predicate ($59$) & $35.0 \rightarrow 70.6$ & $52.5 \rightarrow 78.0$ \\
		No auxiliary predicate ($77$) & $67.1 \rightarrow 92.6$ & $73.2 \rightarrow 93.5$ \\
		\quad auxiliary-predicate gap & $32.1 \rightarrow 22.0$ & $20.7 \rightarrow 15.5$ \\
		\bottomrule
	\end{tabular}
\end{table}

\section{Limitations in Detail}
\label{app:limitations}

This appendix expands the limitations summarized in
Section~\ref{sec:conclusion} and discusses the scope of interactive use.

\emph{Scope of the dialect.} \bench targets the Souffl{\'e} dialect and grades
programs by executing them under Souffl{\'e}. Datalog has no standardized surface
syntax, and engines such as DDlog, Flix, and Scallop differ in typing,
aggregation, and negation, so a program counted as incorrect here may be
well-formed elsewhere. We do not measure how far our findings transfer to other
dialects.

\emph{What the oracle establishes.} Agreement on a finite set of evaluation
variants is not a proof of program equivalence: a wrong program can coincide with
the reference on every variant we test. This limitation is shared with the
text-to-SQL literature and can be bounded but not removed. We bound it by
mutation analysis: the variants distinguish $1{,}440$ of the $1{,}521$ compilable mutants
($94.7\%$), and $13$ of the $70$ survivors are equivalent by an automatic check
(\S\ref{sec:oracle-strength}). Five further operators extend the analysis to negation, aggregates,
grouping, comparison bounds, and constants (Appendix~\ref{app:new-operators}),
but the eight operators together still do not span every error.

\emph{Attribution in the agent setting.} The agents receive execution feedback
and a scaffold, namely a persistent session, file access, and a shell, at the same
time. Without a condition that delivers feedback alone, we cannot tell how much of
their gain comes from each.

\emph{Fixed specifications rather than interactive use.} Our tasks provide a
fixed natural-language question and relation schema. We do not evaluate
incomplete requests, modifications to an existing program, or feedback from a
user. The coding agents receive automated execution feedback on a demonstration
instance, which does not measure their ability to clarify intent with a user.
Our results assess synthesis from a supplied specification, rather than
reliability in interactive programming.

\section{Implications}
\label{sec:discuss}

\noindent \textbf{For LLM development.}
Code-generation benchmarks have repeatedly shown that executable programs are not
necessarily correct programs~\citep{chen2021evaluating,liu2023isurcode,jimenez2024swebench}.
On \bench, compilation and semantic correctness both remain material failure
modes. Declaration checking is still a cheap safeguard, and it accounts for
$5.5$ points of mean Direct EX. The larger opportunity is targeted repair of
recursive rules and decompositions. Agent interaction supplies a strong signal:
the two systems solve $41.0$ and $32.0$ tasks per run that their matched Direct cells miss, most
of the unexplained gains occurring on recursive tasks. Neuro-symbolic systems that combine LLM
sketches with symbolic checking and search are a natural fit for this
domain~\citep{barke2024hysynth}: the model can propose a decomposition, while the
symbolic component enforces declarations, types, safety, and oracle consistency.

\smallskip
\noindent \textbf{For program synthesis research.}
Our results do not suggest that LLMs can replace symbolic Datalog synthesizers.
Existing techniques provide stronger correctness mechanisms through constraints,
syntax-guided search, solver-based rule selection, and provenance-guided
refinement~\citep{albarghouthi2017constraint,si2018syntax,bembenek2023smt2asp,raghothaman2020provenance}.
Their limitation is not semantic rigor but scalability and specification burden.
LLMs offer a complementary strength: they map informal requirements to plausible
rule sketches and predicate decompositions, and from language they reach many
tasks that example-driven search does not (Appendix~\ref{app:symbolic}). A promising
pipeline would therefore separate responsibilities: the LLM proposes candidate
predicates and rules, while a Datalog engine and a symbolic component supply the
declarations and validate safety, type consistency, recursion structure, and
behavior over examples. The division suits \bench because the symbolic side can
remove inexpensive well-formedness failures and localize the residual errors in
recursion and decomposition, where the model's language prior is most useful.

\noindent \textbf{For practitioners.}
Current LLMs can be useful assistants for Datalog programming, but they should
not be treated as autonomous synthesizers. A practical workflow is to use the
model for low-risk drafting tasks: turning an informal query into an initial rule
skeleton, proposing auxiliary predicate names, explaining an existing rule set,
or generating small test cases. The generated program should then be checked with
the same discipline used for hand-written Datalog: compile it with
Souffl{\'e}~\citep{jordan2016souffle}, run it on several representative input
instances, inspect the derived auxiliary predicates, and compare outputs
against an independently constructed oracle. Practitioners should be especially
cautious on tasks involving transitive closure, dataflow propagation, points-to
analysis, negation, arithmetic constraints, or multiple interacting output
relations. These are the settings where a program may look plausible and still
omit a base case, bind the wrong variable, redefine a relation, or derive an
over-approximate result.

LLM output is therefore best read as a starting point for review rather than a
deployable artifact. This workflow allows LLMs to improve developer productivity
while keeping semantic responsibility with executable validation and human
review.

\section{Comparison with Related Benchmarks and Synthesis Settings}
\label{app:benchmark-comparison}

Table~\ref{tab:benchmark-comparison} compares specification interfaces, structural
evaluation, and grading. Symbolic synthesis settings are included to distinguish
their specification and search assumptions from natural-language synthesis.
Entries concern the cited versions; an unreported structural slice does not
imply an absent language capability or weak tests.

\begingroup
\small
\setlength{\tabcolsep}{3pt}
\renewcommand{\arraystretch}{1.0}
\setlength{\defaultaddspace}{1pt}
\setlength{\LTcapwidth}{\textwidth}
\begin{longtable}{@{}>{\raggedright\arraybackslash}p{.16\linewidth}
>{\raggedright\arraybackslash}p{.22\linewidth}
>{\raggedright\arraybackslash}p{.25\linewidth}
>{\raggedright\arraybackslash}p{.32\linewidth}@{}}
\caption{Related benchmark and synthesis settings: specification, structural
evaluation, and grading. Structural coverage does not establish necessity for
every correct implementation.}\label{tab:benchmark-comparison}\\
\toprule
Setting & Specification & Structural evaluation & Grading and test evidence \\
\midrule
\endfirsthead
\caption[]{Related settings (continued).}\\
\toprule
Setting & Specification & Structural evaluation & Grading and test evidence \\
\midrule
\endhead
\midrule
\multicolumn{4}{r}{\textit{Continued on next page}}\\
\endfoot
\bottomrule
\endlastfoot
ALPS\newline\citeyearpar{si2018syntax} & Input facts, membership oracle, metarules &
Recursive Datalog targets; invention within metarule constraints &
Oracle agreement, synthesis time, query count; input-size sensitivity \\
\addlinespace
ProSynth\newline\citeyearpar{raghothaman2020provenance} & Facts, output constraints, candidate rules &
Recursive targets; intermediate schemas supplied in candidate space &
Constraint satisfaction, synthesis time, timeouts \\
\addlinespace
EGS\newline\citeyearpar{thakkar2021example} & Facts and positive/negative output examples &
Nonrecursive relational queries, including union and negation &
Example consistency, manual inspection, synthesis time, unrealizability \\
\addlinespace
GenSynth\newline\citeyearpar{mendelson2021gensynth} & I/O examples and schemas &
Recursion and automatic intermediates; schema restriction described below &
Example fit; held-out Countries F1 \\
\addlinespace
M\"obius\newline\citeyearpar{thakkar2023mobius} & Facts and positive/negative examples &
Recursive targets; invention of predicates with new schemas &
Example fit; unseen-graph accuracy for three tasks \\
\addlinespace
LogicLoc\newline\citeyearpar{xu2026logicloc} & NL queries, schemas, repository context &
Generated Souffl{\'e} rules; recursion and invention not separately evaluated &
Localization accuracy; execution-success ablation; diagnostic mutations guide synthesis \\
\addlinespace
Spider / BIRD\newline\citeyearpar{yu2018spider,li2023can} & NL questions, database schemas; BIRD adds knowledge evidence &
SQL constructs; recursive queries not in reported inventories &
Execution agreement; Spider also set matching, BIRD efficiency; Spider test-suite extension below \\
\addlinespace
Spider 2.0\newline\citeyearpar{lei2025spider2} & NL tasks, databases, documentation, project context &
SQL/Python workflows with CTEs; no recursion-specific slice &
Final-result correctness; evaluator review and syntax-error analysis \\
\addlinespace
HumanEval / EvalPlus\newline\citeyearpar{chen2021evaluating,liu2023isurcode} & Function signatures and NL docstrings &
Python; recursion is not a designated evaluation axis &
Unit tests and pass@$k$; EvalPlus adds reference-based tests and detects additional errors \\
\addlinespace
\bench & NL questions and relation schemas; optional descriptions/examples &
Souffl{\'e}; reference-based recursion and auxiliary-predicate slices &
Held-out output agreement; mutation and additional-input checks; separate CP/EX and failure analysis \\
\end{longtable}
\endgroup

GenSynth introduces intermediates without user-supplied signatures; in the
baseline studied by M\"obius, new schemas are restricted to input/output schemas.
For Spider, \citet{zhong2020semantic} add distilled database test suites selected
to distinguish neighboring SQL queries; this is an evaluation extension, not a
separate benchmark. Held-out accuracy differs from quantitative
test-adequacy evidence; syntax-error categories and execution success differ
from compilation pass rate.

\end{document}